\documentclass[prd,aps,preprint,tightenlines,superscriptaddress,nofootinbib,showpacs,showkeys]{revtex4}
\usepackage{mathrsfs}
\usepackage{amsfonts}
\usepackage{array}
\usepackage{epsfig}
\usepackage{rotating}
\usepackage{multirow}

\usepackage{amsmath}    % need for subequations
\usepackage{verbatim}   % useful for program listings
\usepackage{color}      % use if color is used in text
\usepackage{colordvi}

\usepackage{subfigure}  % use for side-by-side figures
\usepackage{hyperref}   % use for hypertext links, including those to external documents and URLs

\usepackage{pstricks,rotating, graphicx}

\graphicspath{{../pics/}}
\DeclareGraphicsExtensions{.eps,.ps}

\usepackage{pslatex}
\usepackage{txfonts}
\usepackage[latin1]{inputenc}
\usepackage[T1]{fontenc}

\newcommand{\msbar}{$\overline{\mathrm{MS}}\, $}

\begin{document}

\begin{titlepage}
\thispagestyle{empty}
\noindent
DESY 15-161\\
\hfill
August 2015 \\
\vspace{1.0cm}

\begin{center}
  {\bf
    \Large Iso-spin asymmetry of quark distributions and implications \\ for single top-quark production at the LHC\\
  }
  \vspace{1.25cm}
 {\large
   S.~Alekhin$^{\, a,b}$,
   J.~Bl\"umlein$^{\, c}$,
   S.~Moch$^{\, a}$
   and
   R.~Pla\v cakyt\. e$^{\, d}$
   \\
 }
 \vspace{1.25cm}
 {\it
   $^a$ II. Institut f\"ur Theoretische Physik, Universit\"at Hamburg \\
   Luruper Chaussee 149, D--22761 Hamburg, Germany \\
   \vspace{0.2cm}
   $^b$Institute for High Energy Physics \\
   142281 Protvino, Moscow region, Russia\\
   \vspace{0.2cm}
   $^c$Deutsches Elektronensynchrotron DESY \\
   Platanenallee 6, D--15738 Zeuthen, Germany \\
   \vspace{0.2cm}
   $^d$Deutsches Elektronensynchrotron DESY \\
   Notkestra{\ss}e 85, D--22607 Hamburg, Germany \\
 }
  \vspace{1.4cm}
  \large {\bf Abstract}
  \vspace{-0.2cm}
\end{center}
We present an improved determination of the up- and down-quark distributions 
in the proton using recent data on charged lepton asymmetries from $W^\pm$ gauge-boson production at the LHC and Tevatron.
The analysis is performed in the framework of a global fit of parton distribution functions.
The fit results are consistent with a non-zero iso-spin asymmetry of the sea, 
$x(\bar d - \bar u)$, at small values of Bjorken $x\sim 10^{-4}$ 
indicating a delayed onset of the Regge asymptotics of a vanishing $(\bar d - \bar u)$-asymmetry at small-$x$.
We compare with up- and down-quark distributions available in the literature 
and provide accurate predictions for the production of single top-quarks at the LHC, 
a process which can serve as a standard candle for the light quark flavor content of the proton.
\end{titlepage}

\newpage
\setcounter{footnote}{0}
\setcounter{page}{1}

%%
%% ---------------------------------------------------------------------------
%%
\section{Introduction}

The hadro-production of the electroweak gauge bosons $W^\pm$ and $Z$ is an important reaction, 
which can be measured with very high precision in hadron collider experiments,
such as Tevatron or the Large Hadron Collider (LHC). 
The available theoretical predictions match the precision of the experimental data 
thanks to the known radiative corrections to next-to-next-to leading order (NNLO) in QCD 
and to next-to leading order (NLO) in the electroweak sector of the Standard Model. 
The process of $W^\pm$- and $Z$-boson production has, therefore, 
become known as a so-called standard candle process, 
because it provides valuable constraints on the parton content of the colliding protons.
In a global fit of parton distribution functions (PDFs) 
precision data on $W^\pm$- and $Z$-boson production allow to separate 
the quark flavors, in particular up- and down quarks, 
and to disentangle the sea and valence quark PDFs.

Recently, new data sets have become available from the LHC~\cite{Chatrchyan:2013mza,Aaij:2014wba} and the Tevatron
experiment D0~\cite{Abazov:2013rja,D0:2014kma}
based on measurements of the rapidity distributions of charged leptons produced in the decays of the $W^\pm$-bosons 
or charged-lepton pairs from the $Z$-boson decays. 
The kinematic reach to large rapidities makes the data analysis very interesting 
for small-$x$ physics, particularly in the context of the LHC.
In the past, the proton structure at small $x$ has been studied in detail at the HERA collider.
However, the potential for quark flavor separation has always been poor 
due to the underlying deep-inelastic scattering (DIS) process probing dominantly flavor singlet states.
Successful improvements in the determination of the strange quark PDF have recently been undertaken~\cite{Alekhin:2014sya} 
with the help of new neutrino-nucleon DIS data, 
but the non-strange sea at small $x$ could not be separated with existing data thus far.
The latter, therefore, is routinely described by a parameterization with Regge-like asymptotics in global PDF fits.
The new D0 data on electron and muon asymmetries~\cite{Abazov:2013rja,D0:2014kma}
as well as the LHCb data~\cite{Aaij:2014wba} on the measurement of the forward $W^\pm$-boson cross-section in $pp$ collisions
can help to resolve this ambiguity due to the particular kinematics in the forward region.
In addition, the CMS measurement of the muon charge asymmetry~\cite{Chatrchyan:2013mza} assists in the 
improved determination of light parton PDFs due to very high accuracy of the data. 

In this letter we study the impact of the new data on charged lepton
asymmetries from $W^\pm$-boson production and their rapidity distributions   
on the global fit of PDFs by ABM~\cite{Alekhin:2009ni,Alekhin:2012ig,Alekhin:2013nda}. 
Taking the ABM12 fit~\cite{Alekhin:2013nda} based on the world DIS data, 
measurements of Drell-Yan (DY) dimuon production from fixed targets and early LHC data 
on $W^\pm$- and $Z$-boson production
as a reference the impact of individual new data sets is quantified and the resulting shifts
in the up- and down quark PDFs are documented.
The fit results turn out to be consistent with a non-zero iso-spin asymmetry of the sea, 
$I(x) = x(\bar d - \bar u)$, at small values of Bjorken $x\sim 10^{-4}$. 
This indicates a delayed onset of the Regge asymptotics of a vanishing
asymmetry $I(x)$ at small $x$.
Particular care is taken to control potential correlations of the $d$-quark distribution at large $x$ 
and the $\bar d$-quark distribution at small $x$.  
The results are compared with earlier determinations of the asymmetric sea $x(\bar d - \bar u)$ 
at small $x$ obtained in the global fits CT10~\cite{Gao:2013xoa}, CT14~\cite{Dulat:2015mca}, 
MMHT~\cite{Harland-Lang:2014zoa} or the so-called unbiased fit NN3.0~\cite{Ball:2014uwa} 
as well as with the recent {\tt HERAFitter} analysis~\cite{Camarda:2015zba}.

The fit results of the present analysis for the $u$- and $d$-quark distributions are used 
to provide accurate predictions for the production of single top-quarks at the LHC.
This process furnishes a very sensitive test for the light quark flavor content of the proton. 
The respective cross sections are almost complete to NNLO in QCD for $t$-channel single top-quark production~\cite{Brucherseifer:2014ama} 
and measurements at the LHC have been conducted at $\sqrt{s} = 7$~TeV by ATLAS~\cite{Aad:2014fwa}
and at $\sqrt{s} = 8$~TeV by CMS~\cite{Khachatryan:2014iya}.
With the new PDFs of the present fit we obtain updated predictions for the single (anti)top-quark cross sections $\sigma_t$ and $\sigma_{\bar t}$ as well as
their ratio $R_t=\sigma_t/\sigma_{\bar t}$. The latter quantity is particularly
sensitive to PDFs as uncertainties largely cancel in the ratio. 
As a result, we demonstrate in this letter that single top-quark production can serve as a standard candle process at the LHC, cf. \cite{Alekhin:2010dd}.

\section{Set-up of the analysis}
\label{sec:set-up}

The present analysis is based on the update of the ABM12 global fit~\cite{Alekhin:2013nda} 
including recent data on the neutrino-induced dimuon production~\cite{Alekhin:2014sya} and
with the addition of recent data on $W^\pm$- and $Z$-boson production collected by the LHC and Tevatron experiments. 
This new input allows to improve the determination of the quark distributions, 
and, in particular, the separation of the $u$- and $d$-quarks. 
At first instance, these are the LHCb~\cite{Aaij:2014wba} and D0~\cite{D0:2014kma}  
data on forward $W^{\pm}$-boson production providing a constraint on  
the sea quark iso-spin asymmetry $x(\bar{d}-\bar{u})$ in a range of $x$, 
which has not been probed in earlier experiments.
The LHCb statistics of Ref.~\cite{Aaij:2014wba} is collected in the muon 
channel, with the muon pseudo-rapidity $\eta_\mu$ in the range of 
$2 \div 4.5$, which is sensitive to parton momentum fractions $x$ down to $x\sim 10^{-4}$. 
The same applies to the new D0 data on the electron-positron asymmetry~\cite{D0:2014kma} 
covering the electron/positron pseudo-rapidity $\eta_e$ in the range 
of $-3.2 \div 3.2$. The upper margin of this range is smaller than 
one of the LHCb data~\cite{Aaij:2014wba}, 
however due to the difference in 
the collision energy of Tevatron and the LHC, the D0 data probe the
quark distributions down to $x\sim 10^{-4}$ as well as the LHCb ones. 
The forward kinematics of the LHCb and D0 experiments  
is also sensitive to the quark distributions at large $x$ and, in 
particular, can be used to pin down the $d$-quark distribution, 
which is  poorly known in this region. 
The final D0 data on the muon charge asymmetry~\cite{Abazov:2013rja} 
span the region of $| \eta_{\mu}|<2$. 
Therefore they put a constraint on the quark iso-spin asymmetry mostly 
at $x\gtrsim 0.001$. 
The updated CMS data on the muon charge asymmetry~\cite{Chatrchyan:2013mza} with $|\eta_{\mu}|<2.4$ 
are even more limited in the parton momentum coverage. 
However, they provide important experimental input due to unprecedented accuracy. 
We also add to the analysis recent data on the $Z$-boson 
production in the forward region~\cite{Aaij:2015gna,Aaij:2015vua} 
obtained by the LHCb experiment at
the collision energies of 7 and 8 TeV. These data provide a 
complementary constraint on the PDF combination and allow further improvements in
disentangling the quark species at small and large $x$. 
On the other hand, earlier data from CMS 
on the electron charge asymmetry~\cite{Chatrchyan:2012xt} 
and from LHCb on forward $W^\pm$-boson production~\cite{Aaij:2012vn} 
are removed from the fit in view of their low statistical significance 
as compared to the more recent data of these experiments.
The deuteron data by the NMC, BCDMS, and SLAC experiments employed in the 
ABM12 fit are also not included into the present analysis in order 
not to be exposed to uncertainties related to modeling of the nuclear effects. 

\begin{table}[ht!]
\renewcommand{\arraystretch}{1.3}
\begin{center}                   
{\small                          
\begin{tabular}{|c|c|c|c|c|c|c|c|}   
\hline                           
\multicolumn{2}{|c|}{Experiment}                      
&ATLAS
&CMS  
&\multicolumn{2}{c|}{D0}
&\multicolumn{2}{c|}{LHCb}
\\
\hline                                                    
\multicolumn{2}{|c|}{$\sqrt s$~(TeV)}                      
&7                         
&7  
&\multicolumn{2}{c|}{1.96}
&{7}
&8                                                        
\\                                                        
\hline
\multicolumn{2}{|c|}{Final states} 
& $W^+\rightarrow l^+\nu$
& $W^+\rightarrow \mu^+\nu$
&$W^+\rightarrow \mu^+\nu$
& $W^+\rightarrow e^+\nu$
&$W^+\rightarrow \mu^+\nu$
& $Z\rightarrow e^+e^-$                                                        
\\
\multicolumn{2}{|c|}{ }                                          
& $W^-\rightarrow l^-\nu$
&$W^-\rightarrow \mu^-\nu$
&$W^-\rightarrow \mu^-\nu$
&$W^-\rightarrow e^-\nu$
&$W^-\rightarrow \mu^-\nu$
&                                                         
\\                                                        
\multicolumn{2}{|c|}{ }                                          
& $Z\rightarrow l^+l^-$
&                                                         
&
&                                                         
& $Z\rightarrow \mu^+\mu^-$
&                                                         
\\                                                        
\hline                           
\multicolumn{2}{|c|}{Reference}                      
&\cite{Aad:2011dm}                         
&\cite{Chatrchyan:2013mza}
&\cite{Abazov:2013rja}
&\cite{D0:2014kma}
&\cite{Aaij:2015gna}
&\cite{Aaij:2015vua}
\\
\hline                                                    
\multicolumn{2}{|c|}{Cut on the lepton $P_T$ }                      
&$P_T^l>20~{\rm GeV}$      
&$P_T^{\mu}>25~{\rm GeV}$   
&$P_T^{\mu}>25~{\rm GeV}$
&$P_T^{e}>25~{\rm GeV}$
&$P_T^{\mu}>25~{\rm GeV}$
&$P_T^{e}>20~{\rm GeV}$                       
\\
\hline                                                    
\multicolumn{2}{|c|}{Luminosity (1/fb)}                      
&0.035                         
&4.7  
&7.3
&9.7
&1.
&2.                        
\\                                                        
\hline                                                    
\multicolumn{2}{|c|}{$NDP$}
&30                      
&11  
&10
&13
&31
&17                            
\\                                                        
\hline
\end{tabular}
}
\caption{
\label{tab:data-inp}
\small 
The different data samples of $W^\pm$- and $Z$-boson production from 
LHC and Tevatron used in the present analysis.} 
\end{center}
\end{table}
The collider data on $W^\pm$- and $Z$-boson production 
used in the present analysis are summarized in Table~\ref{tab:data-inp}. 
This set is essentially different from the respective data used in the ABM12 fit. 
Indeed, many high-statistical samples are added now, 
while only the ATLAS data on $W^\pm$- and $Z$-boson production~\cite{Aad:2011dm} 
and the LHCb data on $Z$-boson production in the electron channel~\cite{Aaij:2012mda} are used in both cases.

In order to take advantage of this input to constrain the PDFs 
fully exclusive QCD calculations of the $W^\pm$-boson production process are required,
because the data sets of Table~\ref{tab:data-inp} are obtained in a restricted phase space 
due to limited detector acceptance and the $W^\pm/Z$-boson event selection criteria
with a cut on the lepton transverse momentum $p_T^l$ imposed. 
Fully differential QCD predictions for lepton rapidity distributions  
are implemented up to NNLO in the publicly available code {\tt FEWZ} (version 3.1)~\cite{Li:2012wna,Gavin:2012sy}. 
{\tt FEWZ} (version 3.1) provides a convenient capability to estimate
uncertainties in the cross sections due to the PDFs.
The data sets newly included into our analysis are compared with the
NNLO {\tt FEWZ} predictions based on the ABM12, CT10, CT14, MMHT, and NN3.0 PDFs 
in Figs.~\ref{fig:tev}, \ref{fig:lhcb}, \ref{fig:lhc} and \ref{fig:lhcbz}. 

\begin{figure}[tbh]
\centerline{
  \includegraphics[width=15.0cm]{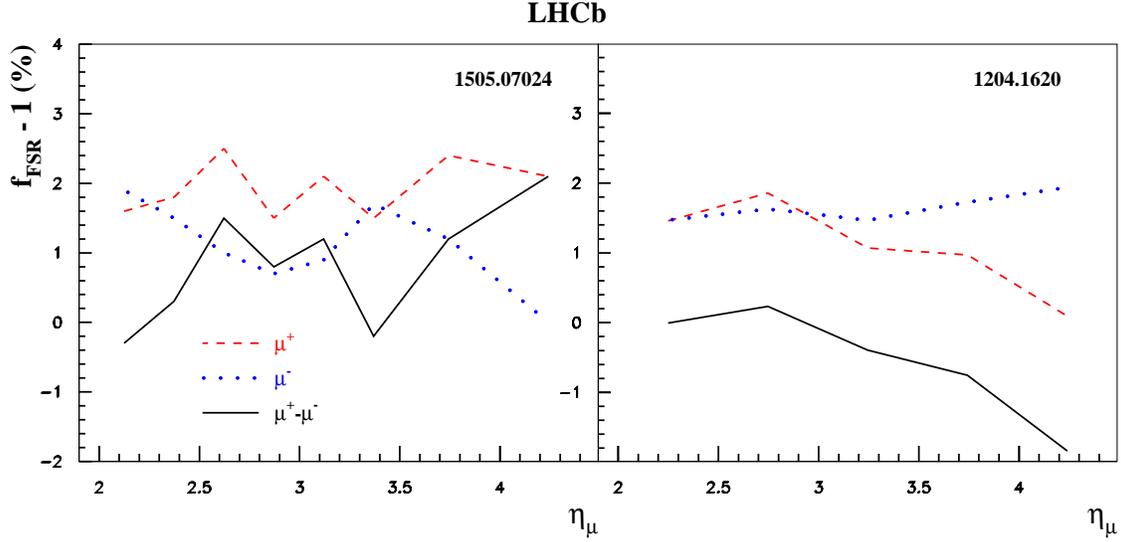}}
  \caption{\small
    \label{fig:fsr}
    Left panel: Correction for the final-state-radiation (FSR) effects applied 
    in the LHCb analysis~\cite{Aaij:2015gna} to the data on the 
    pseudo-rapidity distribution of muons produced in forward $W^\pm$-decays
    (dashes: $\mu^+$, 
    dots: $\mu^-$, solid: difference between these two). Right panel: The same 
    for the earlier LHCb analysis of Ref.~\cite{Aaij:2012vn}.
}
\end{figure}

The predictions agree in general with the data~\footnote{The LHCb data of Ref.~\cite{Aaij:2014wba} on 
$W^{\pm}$-boson production undershoot the predictions at $\eta_{\mu}\sim 2$. 
However they also come in agreement once the trigger efficiency is 
fixed~\cite{Aaij:2015gna}, cf. Fig.~\ref{fig:lhcb}.}. 
However, in places random fluctuations beyond the uncertainties appear. 
In particular, those fluctuations are observed for the CMS and LHCb data
on the lepton asymmetry, cf. Fig.~\ref{fig:lhc}. 
This prevents to obtain an ideal value of $\chi^2$ for these two samples. 
In the LHCb sample one specific data point at $\eta_{\mu}=3.375$ deviates from the general trend most significantly. 
It is worth noting that the shape of final state radiation (FSR) corrections 
to account for QED corrections in $W^\pm$- and $Z$-boson decays applied 
to the data of Ref.~\cite{Aaij:2015gna} demonstrates an anomalous irregularity 
precisely at this value of $\eta_{\mu}$, cf. Fig.~\ref{fig:fsr}. 
In contrast, in the earlier LHCb analysis of the muon asymmetry~\cite{Aaij:2012vn}
the FSR corrections employed exhibit a smooth shape and for such a shape
the fluctuation in the pseudo-rapidity distribution at $\eta_{\mu}=3.375$ vanishes.  
The treatment of FSR in LHCb differs 
between the respective analysis.
In Ref.~\cite{Aaij:2014wba} the FSR corrections 
are computed with the {\tt PHOTOS} Monte Carlo~\cite{Golonka:2005pn}
while Ref.~\cite{Aaij:2015gna} has determined FSR corrections 
from the mean of simulations done with {\tt Herwig++}~\cite{Bahr:2008pv} and
{\tt Pythia8}~\cite{Sjostrand:2007gs}. 
Having no possibility to validate this issue, 
we simply drop the problematic points at $\eta_{\mu}=3.375$ 
in the LHCb $W^\pm$-boson data sample from the fit. 
This reduces the value of $\chi^2$ by some 10 units 
and brings it down to a reasonable level, 
although the FSR corrections still may need clarification for the overall
LHCb kinematics of Ref.~\cite{Aaij:2015gna} and other publications of this experiment. 
In particular, this concerns the uncertainty in the FSR correction, which has been recently
estimated as $\pm 0.5\%$, which is much bigger than the value of $\pm 0.2\%$ quoted in the earlier LHCb 
analysis of Ref.~\cite{Aaij:2012vn}. 

In line with this revision we also expand the FSR correction uncertainty in the 
LHCb $Z$-boson production data of Ref.~\cite{Aaij:2012mda} to the same value of 
$\pm 0.5\%$. However, the early LHCb data of Ref.~\cite{Aaij:2012mda} are anyway
in a certain tension with more recent LHCb $Z$-boson production samples collected 
in different decay channels and at different collision energies. 
Indeed, the electron data of Ref.~\cite{Aaij:2012mda} 
collected at the center of mass (c.m.s.) energy $\sqrt{s}=7~{\rm TeV}$ 
overshoot (undershoot) the muon ones of Ref.~\cite{Aaij:2015vua} at the same $\sqrt{s}$ 
when the pseudo-rapidity is bigger (smaller) than $\eta\sim 3$, 
while the trend of LHCb electron data~\cite{Aaij:2015gna} 
collected at $\sqrt{s}=8~{\rm TeV}$ is similar to the one of Ref.~\cite{Aaij:2015vua}, cf. Fig.~\ref{fig:lhcb}. 

\begin{table}[th!]
\renewcommand{\arraystretch}{1.3}
\begin{center}                   
{\small                          
\begin{tabular}{|c|c|c|c|}                                                    
\hline                           
{Experiment}                      
&\multicolumn{3}{c|}{LHCb}
\\
\hline                           
{$\sqrt s$~(TeV)}                      
&\multicolumn{2}{|c|}{7}
&8                                                        
\\                                                        
\hline
{Final state} 
& $Z\rightarrow e^+e^-$                                                        
&$Z\rightarrow \mu^+\mu^-$
& $Z\rightarrow e^+e^-$                                                        
\\
\hline                           
{Reference}                      
&\cite{Aaij:2012mda}
&\cite{Aaij:2015gna}
&\cite{Aaij:2015vua}
\\                                                        
\hline                                                    
{$NDP$}
&9
&17
&17                            
\\                                                        
\hline
{$\chi^2_1$}
&9.4
&--
&--
\\                                                        
\hline
{$\chi^2_2$}
&--
&11.3
&--
\\
\hline
{$\chi^2_3$}
&--
&--
&$19.1$
\\
\hline
{$\chi^2_4$}
&16.4
&14.8
&--
\\
\hline
{$\chi^2_5$}
&16.2
&--
&21.4
\\
\hline
{$\chi^2_6$}
&--
&10.5
&18.3
\\
\hline                                          
\end{tabular}
}
\caption{
\label{tab:lhcb}
\small 
The values of $\chi^2_i$ obtained in the present analysis
for the LHCb data on $Z$-boson production~\cite{Aaij:2012mda,Aaij:2015gna,Aaij:2015vua}, 
where index $i=1\dots 6$ denotes variants of the fit based on different combinations of those data. 
} 
\end{center}
\end{table}
This tension is quantified by the value of $\chi^2$ obtained in variants of our analysis, 
which include different combinations of the LHCb $Z$-boson data 
without adding the other data sets listed in Table~\ref{tab:data-inp} 
in order to provide a clean comparison.
A good value of $\chi^2$ is only achieved 
when one single LHCb $Z$-boson data set is included into the fit (variants \#1,2,3 of Table~\ref{tab:lhcb}). 
A slightly increased value of $\chi^2$ appears only for the electron sample at $\sqrt{s}=8$~TeV 
demonstrating fluctuations beyond the uncorrelated errors (cf. Fig.~\ref{fig:lhcb}). 
However, upon combination of the electron sample at $\sqrt{s}=7$~TeV with the 
muon one at $\sqrt{s}=7$~TeV and the electron one at $\sqrt{s}=8$~TeV (variants \#4,5)
the value of $\chi^2$ increases substantially. 
For the combination of muon data at $\sqrt{s}=7$~TeV and the electron ones 
at $\sqrt{s}=8$~TeV samples (variant \#6) on the other hand the value of
$\chi^2$ remains essentially the same as for the fits with single samples.
As a consequence of this observation we exclude the LHCb electron sample~\cite{Aaij:2012mda} 
at $\sqrt{s}=7$~TeV from our analysis. 
Moreover, the errors in those data are also substantially bigger 
than in the ones of Refs.~\cite{Aaij:2015vua,Aaij:2015gna}. 

The theoretical predictions displayed in Figs.~\ref{fig:tev}, \ref{fig:lhcb}, \ref{fig:lhc} and \ref{fig:lhcbz}
demonstrate some random fluctuations due to the numerical integration of the
NNLO QCD matrix elements in {\tt FEWZ}. 
These fluctuations are inessential for the purpose of qualitative comparison. 
However, they cause an additional contribution to the value of $\chi^2$ 
when quantifying the agreement of those theory predictions with the data. 
To suppress this contribution to a marginal level one has to provide 
an integration accuracy better than the data uncertainties, i.e., 
at the level of several \textperthousand\, for the high-statistics 
$W^\pm$- and $Z$-boson data samples used in our analysis. 
To achieve such an accuracy one has to generate $2\cdot10^{9}$ {\tt FEWZ} 
histories for each bin in pseudo-rapidity distributions.

\begin{figure}[tbh]
\centerline{
  \includegraphics[width=9cm]{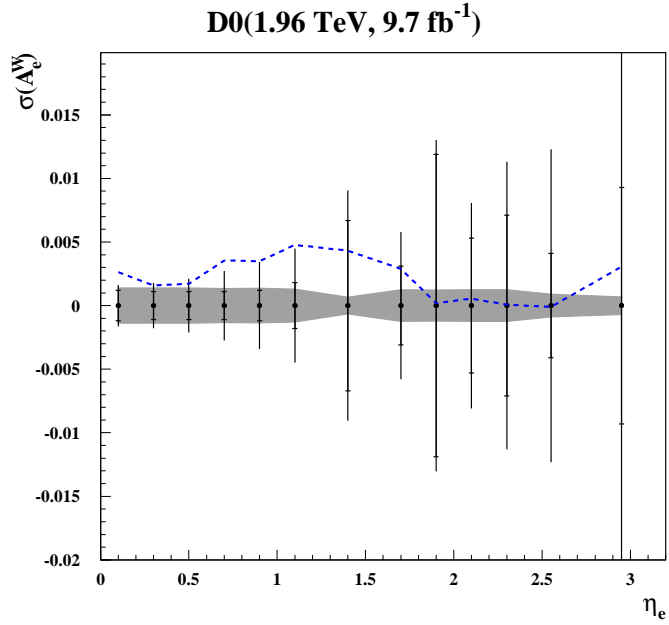}}
  \caption{\small
    \label{fig:d0eacc}
    The experimental uncertainties in the D0 data on the electron 
    asymmetry~\cite{D0:2014kma} (points) compared with 
    the numerical integration accuracy achieved with 
    {\tt FEWZ} (version 3.1)~\cite{Li:2012wna,Gavin:2012sy}
    in the generation of the NNLO grids employed in present analysis
    (shaded area). 
    The impact of the NNLO corrections on the 
    hard scattering coefficient functions is indicated by the dashed curve.
}
\end{figure}

The results obtained in this way are illustrated in Fig.~\ref{fig:d0eacc} 
on the example of the D0 data on electron-positron asymmetry~\cite{D0:2014kma}.  
The integration accuracy displayed in Fig.~\ref{fig:d0eacc} is smaller than the 
experimental uncertainties, particularly in the forward region, 
where the data are less accurate. 
At this rate the value of $\chi^2$ is only marginally sensitive 
to the integration uncertainties in the QCD matrix elements 
and the generation of additional {\tt FEWZ} histories is not necessary.
However, the typical CPU time necessary to accomplish these computations is quite
substantial, of the order ${\cal O}(10^{4}{\rm h})$ for each data set of Table~\ref{tab:data-inp}. 
Therefore they cannot be repeated iteratively, with variations of the PDFs in the fit. Instead, we compute the prediction only once, 
but for each eigenvector encoding the uncertainties in an initial PDF set. 
Afterwards, upon varying the PDFs in a fit, we obtain the  
$W^\pm$- and $Z$-boson cross sections corresponding to the current PDF values 
as an interpolation between the entries of this eigenvector prediction grid
(cf. Ref.~\cite{Alekhin:2013nda} for more details). 
Conceptually, this approach is based on the algorithm employed in the code 
{\tt fastNLO} for the fast computation of differential distributions 
of jets in hadro-production with account of the NLO QCD corrections~\cite{Kluge:2006xs}. 
The main difference with {\tt fastNLO} is the selection of the PDF basis 
used to generate the prediction grid. 
In our case it is much smaller than the one of {\tt fastNLO}. 
However it is sufficient once the PDF variations due to new data included 
are within their uncertainties. 
As an advantage, the computing time necessary for the grid generation
is greatly reduced as compared to the case of a general PDF basis employed in {\tt fastNLO}. 

To meet the increased statistical significance of the new data sets used, 
cf. Table~\ref{tab:data-inp}, we allow more flexibility in the fitted PDFs 
by releasing a previously imposed constraint on the sea quark iso-spin asymmetry 
\begin{equation}
I(x)=x(\bar{d}_s-\bar{u}_s) \sim x^{\delta_{ud}},
\label{eq:iso}
\end{equation}  
where $\delta_{ud}\sim 0.5$ due to Regge model arguments. 
In the present analysis the boundary conditions 
for the $u$- and $d$ sea quark distributions 
at the scale of $\mu_0^2$ are chosen
\begin{equation}
xu_s(x,\mu_0^2)=\bar{u}_s(x,\mu_0^2) =A_{us}(1-x)^{b_{us}}x^{a_{us}P_{us}(x)}\, ,
\label{eq:usea}
\end{equation}
\begin{equation}
xd_s(x,\mu_0^2)=\bar{d}_s(x,\mu_0^2) =A_{ds}(1-x)^{b_{ds}}x^{a_{ds}P_{ds}(x)}\, ,
\label{eq:dsea}
\end{equation}  
respectively, where 
\begin{equation}
P_{us}(x)=(1+\gamma_{-1,us}\ln x)(1+\gamma_{1,us}x)\, ,
\end{equation}
\begin{equation}
P_{ds}(x)=(1+\gamma_{-1,ds}\ln x)(1+\gamma_{1,ds}x)\, .
\end{equation}
In this representation the parameters 
$a_{us},a_{ds}$ and $\gamma_{-1,us},\gamma_{-1,ds}$ control the low-$x$ PDF behavior. 
Since they are fitted independently, as well as the normalization parameters
$A_{us}, A_{ds}$, the small-$x$ shape of the sea quark iso-spin asymmetry in the parameterization 
Eqs.~(\ref{eq:usea}), (\ref{eq:dsea}) is unconstrained. 
The remaining PDF shapes for the valence, strange, and gluon distributions used in the present analysis 
are the same as the ones of the ABM12 fit. 

\begin{figure}[tbh]
\centerline{
  \includegraphics[width=15.0cm]{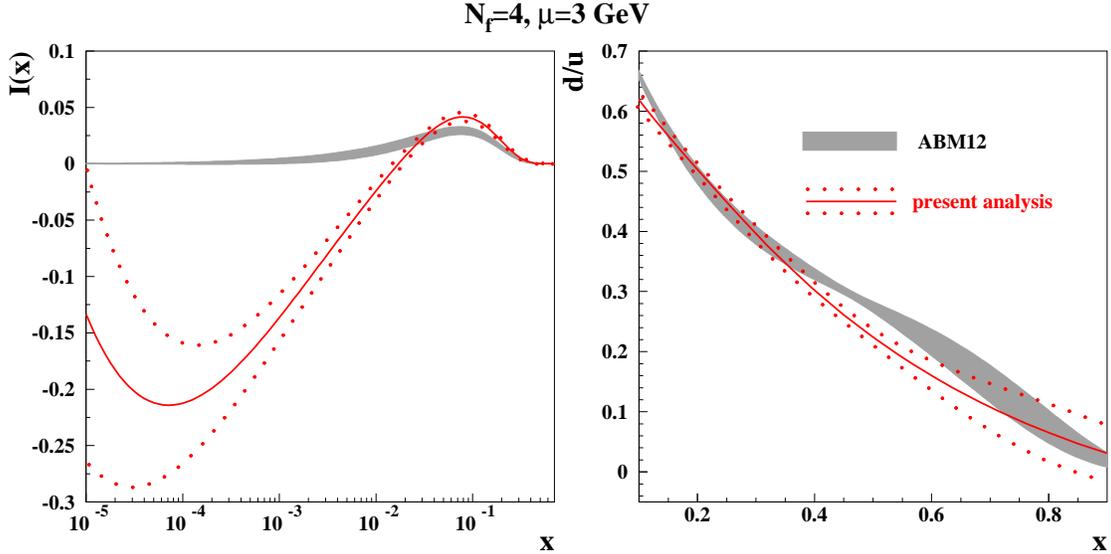}}
  \caption{\small
    \label{fig:udm}
    Left: The iso-spin asymmetry of the sea $I(x)$ 
    in the 4-flavor scheme at the factorization scale $\mu=3~{\rm GeV}$
    obtained in the present variant of the ABM12 analysis 
    (solid lines: central value, dots: $1\sigma$ error band) 
    in comparison to the $1\sigma$ band of $I(x)$
    obtained in the ABM12 
    fit~\cite{Alekhin:2013nda} (shaded area).
    Right: Same for the ratio $d/u$ at large $x$. 
   }
\end{figure}

Having checked the statistical significance of the data sets in Table~\ref{tab:data-inp} 
we find that the fitted value of the parameter $\gamma_{-1,ds}$ is comparable to 0 within uncertainties. 
Furthermore, the value of $\chi^2$ remains essentially the same when it is fixed at 0. 
Therefore, we impose the constraint $\gamma_{-1,ds}=0$ for all variants of the fit considered.
The sea quark iso-spin asymmetry at the initial scale of $\mu_0=3~{\rm GeV}$ obtained in this way 
is displayed in Fig.~\ref{fig:udm} in comparison with the earlier ABM12 determination. 
Due to the relaxed functional form in Eqs.~(\ref{eq:usea}), (\ref{eq:dsea}) 
its central value goes lower than the ABM12 one at $x \lesssim 0.01$. 
This range of kinematics is mostly controlled by the forward $W^\pm$-boson data 
of D0~\cite{D0:2014kma} and LHCb~\cite{Aaij:2014wba}, 
which clearly prefer a negative value of the asymmetry at $x\sim 10^{-3}$. 
Meanwhile the central $W^\pm$- and $Z$-boson data obtained at the LHC~\cite{Aad:2011dm,Chatrchyan:2013mza} 
demonstrate the same trend, although with less statistical significance (cf. Fig.~\ref{fig:udm}). 
In this context it is worth noting that the description of the ATLAS data 
in the present analysis is improved as compared to the ABM12 one 
(cf. Table~\ref{tab:data-chi} in Section~\ref{sec:comp}), 
evidently due to the relaxed shape of the sea quark distributions at small $x$. 
At $x\lesssim 10^{-4}$ the shape of $I(x)$ demonstrates turnover. 
Therefore, the Regge asymptotics still may recover at even smaller values of $x$, say $x\sim 10^{-6}$. 
However, this observation is based on an unsophisticated extrapolation of Eqs.~(\ref{eq:usea}), (\ref{eq:dsea}) 
to those kinematics, which is not controlled by existing data. 

The forward $W^\pm$-boson data also control quark distributions at large $x$ and,
in particular, allow to constrain the $d/u$ ratio in the absence 
of deuteron-target DIS data employed for this purpose in the ABM12 fit. 
The shape of the $d/u$ ratio determined in the present analysis is in agreement 
with the ABM12 result (cf. Fig.~\ref{fig:udm}). 
This confirms the validity of the Kulagin-Petti approach~\cite{Kulagin:2004ie} used for modeling 
nuclear effects in the ABM12 fit. 
A general form of the boundary conditions 
for the valence $u$- and $d$-quarks defining the $d/u$ ratio at large $x$ is the same
as for the sea quarks and reads
\begin{equation}
xu_v(x,\mu_0^2) =A_{uv}(1-x)^{b_{uv}}x^{P_{uv}(x)}\, ,
\label{eq:uv}
\end{equation}
\begin{equation}
xd_v(x,\mu_0^2)=A_{dv}(1-x)^{b_{dv}}x^{P_{dv}(x)}\, ,
\label{eq:dv}
\end{equation}  
respectively. However, in this case the polynomials $P_{uv}(x)$ and $P_{dv}(x)$ read
\begin{equation}
P_{uv}(x)=a_{uv}+\gamma_{1,uv}x+\gamma_{2,uv}x^2+\gamma_{3,uv}x^3\, ,
\end{equation}
and 
\begin{equation}
P_{dv}(x)=a_{dv}+\gamma_{1,dv}x+\gamma_{2,dv}x^2+\gamma_{3,dv}x^3\, ,
\end{equation}
respectively, allowing more flexibility for the large-$x$ PDF shape. 
The coefficient $\gamma_{3,dv}$ 
could not be constrained by the data used in the ABM12 fit, therefore it was set to 0. 
The fitted value of $\gamma_{3,dv}$ obtained using the data of Table~\ref{tab:data-inp} is 
also comparable with 0 within uncertainties. 
So $\gamma_{3,dv}$ is set to 0 in the present analysis as well. 
The potential of the collider data in the determination of the $d/u$ ratio at large $x$ is 
comparable to the one of the deuteron data used in the ABM12 fit. 
The D0 electron data demonstrate the best sensitivity to $d/u$ preferring 
a somewhat lower value as compared with the LHCb ones,
cf. Fig.~\ref{fig:udm}. 
It is worth noting that the combination of the proton-target DIS 
and $Z$-boson production data also provide some constraint on the $d/u$ ratio 
at large $x$ (cf. the shaded area in the right panel of Fig.~\ref{fig:udm}). 
Such a determination is consistent with the results of our combined analysis, 
although being much less accurate. 

\begin{table}[th!]
\renewcommand{\arraystretch}{1.3}
\begin{center}                   
{\small                          
\begin{tabular}{|c|c|c|}                                                    
\hline                           
{Experiment}                      
&{D0}
&{LHCb}
\\
\hline                           
{$\sqrt s$~(TeV)}
&1.96                      
&{7}                                   
\\                                                        
\hline
{Final state} 
&  $W^+\rightarrow e^+\nu$                                    
&$W^+\rightarrow e^+\nu$                                                      
\\                                                        
&  $W^-\rightarrow e^-\nu$                                    
& $W^-\rightarrow e^-\nu$ 
\\
\hline                           
{Reference}                      
&\cite{D0:2014kma}
&\cite{Aaij:2015gna}
\\                                                        
\hline                                                    
{$NDP$}
&13
&14                            
\\                                                        
\hline
{$\chi^2_1$}
&9.5
&--
\\                                                        
\hline
{$\chi^2_2$}
&--
&12.4
\\
\hline
{$\chi^2_3$}
&16.1
&19.6
\\
\hline                                          
\end{tabular}
}
\caption{
\label{tab:lhcbd0}
\small 
Same as Table~\ref{tab:lhcb} for combinations of the D0~\cite{D0:2014kma} and LHCb~\cite{Aaij:2015gna} data.
} 
\end{center}
\end{table}
The overall quality of the collider data description achieved in our analysis is 
sufficiently good, cf. Table~\ref{tab:data-chi} in Section~\ref{sec:comp}. 
However, in cases we observe a certain deterioration of the $\chi^2$ values 
as compared to the variants of fit with only one single $W^\pm$- and $Z$-boson
data set employed in combination with the world DIS and the 
fixed-target DY dimuon data used in the ABM12 fit. 
This indicates a tension between the data of certain experiments considered. 
To the most extent, this appears for the LHCb and D0 data sets as follows from 
a comparison of the $\chi^2$ values obtained in variants of our analysis 
based on including different combinations of the LHCb and D0 data 
of Refs.~\cite{Aaij:2015gna,Aaij:2015vua,D0:2014kma}
while the other collider data of Table~\ref{tab:data-chi} are omitted, 
in analogy to the study of mutual consistency of the $Z$-boson LHCb data above (cf. Table~\ref{tab:lhcb}). 
The value of $\chi^2$ increases significantly upon combining the D0 data 
on the $W^\pm$-boson production~\cite{D0:2014kma} with the LHCb ones~\cite{Aaij:2015gna} (variant \#3 
of Table~\ref{tab:lhcbd0})\footnote{The problematic data points of 
LHCb~\cite{Aaij:2015gna} corresponding to the muon pseudo-rapidity $\eta_\mu=3.375$ have been not used in the comparisons
of Table~\ref{tab:lhcbd0}, as we have discussed above.}. 
Additional methodical studies are still necessary in order to resolve the
tension observed either in favor of the D0 or of LHCb experiment. 
However, in any case, the discrepancies are at the level of 1$\sigma$ fluctuations in the value of $\chi^2$. 
Therefore the PDF shapes preferred by each experiment are basically comparable within the uncertainties.

\section {Comparison with other determinations}
\label{sec:comp}

\begin{figure}[tbh]
\centerline{
  \includegraphics[width=9cm]{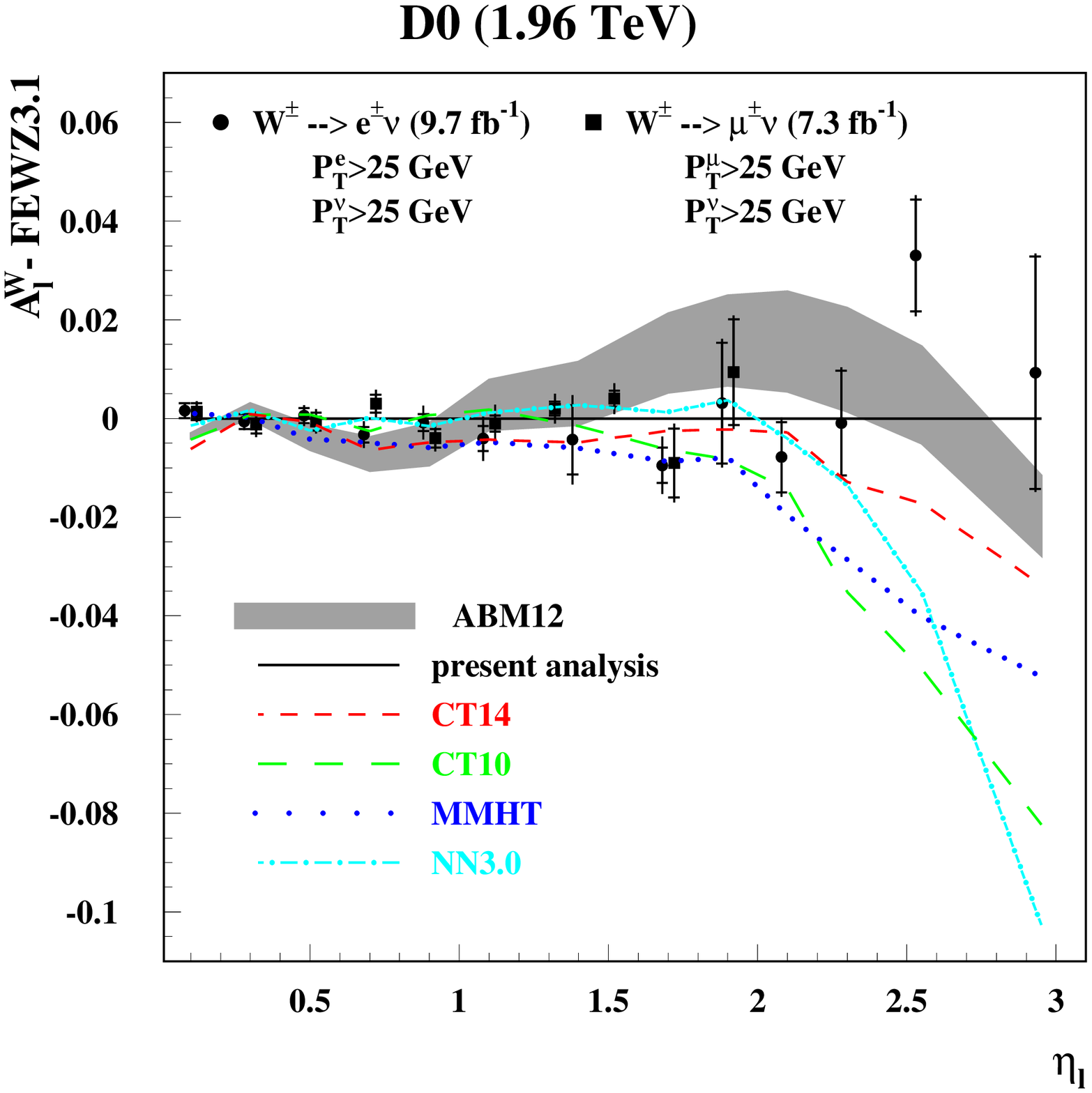}}
  \caption{\small
    \label{fig:tev}
    The pulls of the new D0 data on the charged lepton asymmetry for $p\bar{p} \to W^\pm+X \to l^\pm \nu$
    for electrons~\cite{D0:2014kma} (circles) and muons~\cite{Abazov:2013rja} (squares)
    at the Tevatron  with $\sqrt s = 1.96$~TeV 
    as a function of the pseudo-lepton rapidity $\eta_l$ 
    normalized to the predictions with {\tt FEWZ} (version 3.1)
    at NNLO in QCD computed in the present analysis.
    For comparison, the results of the ABM12 fit
    with the corresponding 1$\sigma$ band of the combined PDF+$\alpha_s$ uncertainty and the central fits of CT10
    CT14, MMHT, and NN3.0 are displayed.
}
\end{figure}

\begin{figure}[tbh]
\centerline{
  \includegraphics[width=15.0cm]{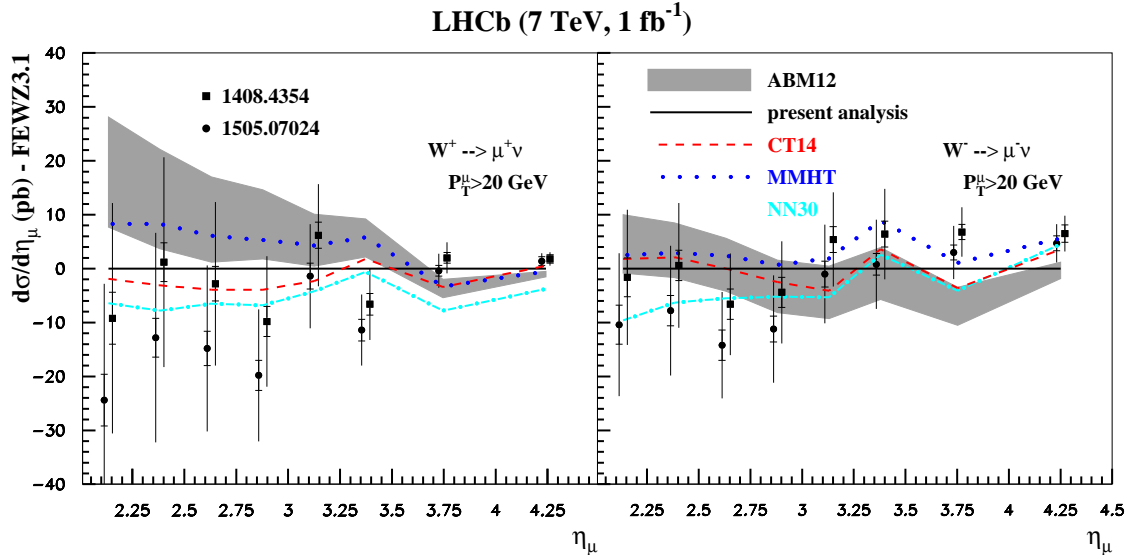}}
  \caption{\small
    \label{fig:lhcb}
    Same as Fig.~\ref{fig:tev} for the LHCb data~\cite{Aaij:2015gna} on 
    the production of $W^+$- (left) and $W^-$-bosons (right) in the $pp$ collision  
    at $\sqrt s = 7$~TeV (circles) in comparison with those data 
    before a trigger efficiency has been fixed~\cite{Aaij:2014wba} 
    (squares). 
}
\end{figure}

\begin{figure}[tbh]
\centerline{
  \includegraphics[width=15.0cm]{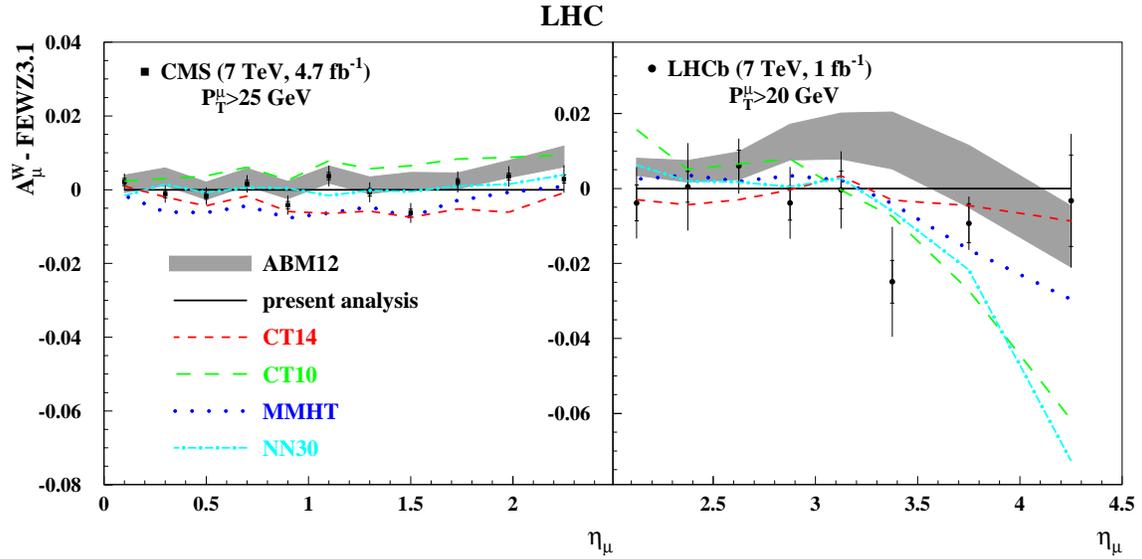}}
  \caption{\small
    \label{fig:lhc}
    Same as Fig.~\ref{fig:tev} for the new CMS~\cite{Chatrchyan:2013mza} 
    and LHCb data \cite{Aaij:2014wba} on the muon charge asymmetry in inclusive
    $pp \to W^\pm+X \to \mu^\pm \nu$ production at the LHC with $\sqrt s = 7$~TeV.
}
\end{figure}

\begin{figure}[tbh]
\centerline{
  \includegraphics[width=15.0cm]{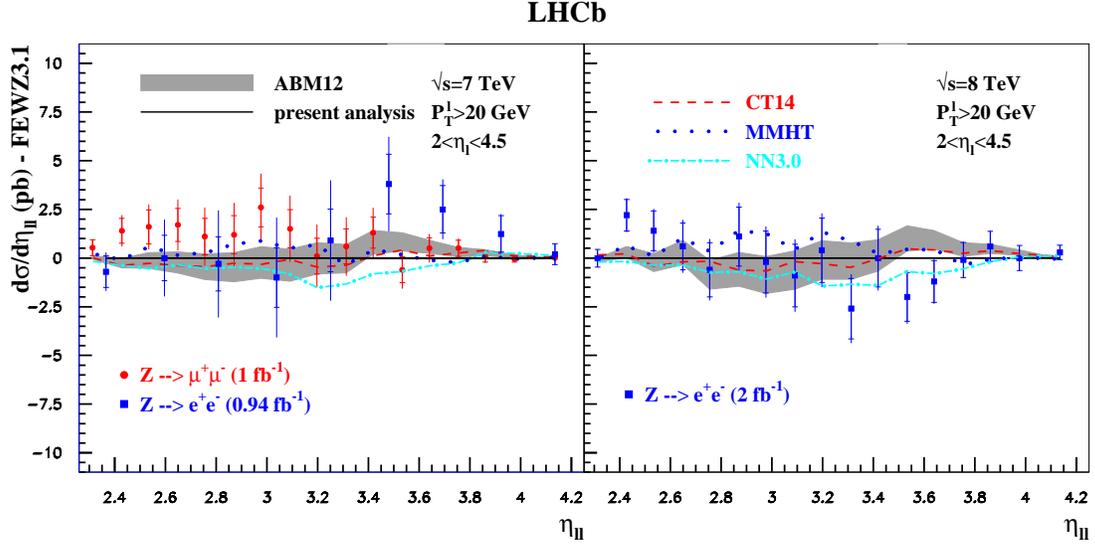}}
  \caption{\small
    \label{fig:lhcbz}
    Same as Fig.~\ref{fig:tev} for the LHCb data on the $Z$-boson 
    production~\cite{Aaij:2012mda,Aaij:2015gna,Aaij:2015vua}
    at the c.m.s. energy of 7~TeV (left) and 8~TeV (right) 
    in the muon channel (circles) and the electron channel (squares) in comparison 
    to the ABM12 predictions. 
}
\end{figure}

\begin{figure}[tbh]
\centerline{
  \includegraphics[width=9cm]{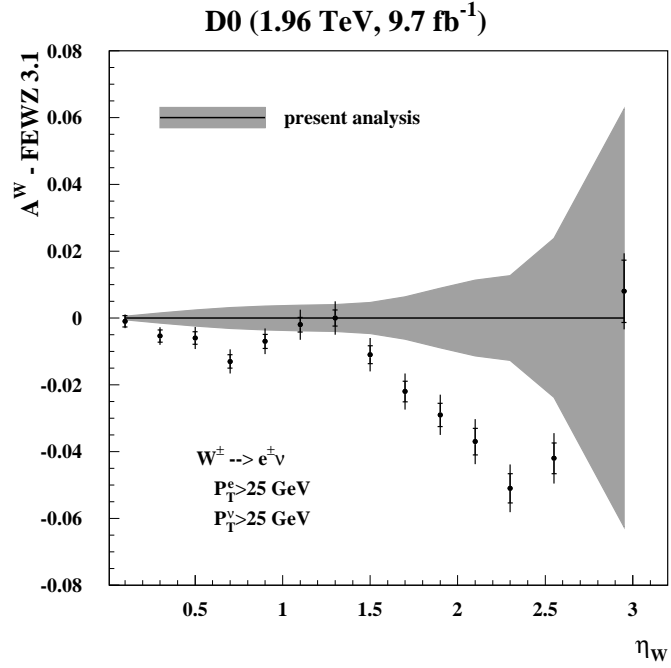}}
  \caption{\small
    \label{fig:d0w}
Same as Fig.~\ref{fig:tev} for the D0 data on the charged 
$W^\pm$-asymmetry~\cite{Abazov:2013dsa} extracted from the D0 data 
on the electron charge asymmetry~\cite{D0:2014kma} as a function of the $W^\pm$-rapidity $\eta_W$. 
The shaded area displays the PDF uncertainties in the predictions. }
\end{figure}

Constraints on the iso-spin asymmetry in quark distributions coming from 
the D0 Tevatron data have also been recently considered in the {\tt HERAFitter} 
analysis~\cite{Camarda:2015zba}. However, in contrast to our case, major parts
of the results in Ref.~\cite{Camarda:2015zba} are based on the data 
for the charge asymmetry of $W^\pm$-bosons~\cite{Abazov:2013dsa}, rather than on the lepton asymmetry~\cite{Abazov:2013dsa}. 
In fact, these two sets stem from the same data sample and the $W^\pm$-asymmetry is reconstructed from the 
lepton distributions by solving for the kinematics of the $W^\pm$-decay with the 
world-average constraint on the $W^\pm$-mass value. 
This solution is not unique and, therefore, 
the final $W^\pm$-distribution is obtained in a statistical way, 
by modeling probabilities of those solutions. 
The D0 modeling is based on the early CTEQ PDFs, 
which do not agree with ours in the relevant range of kinematics for the data of  
Ref.~\cite{Abazov:2013dsa}, cf. Fig.~\ref{fig:tev}. 
This difference does not allow a consistent use of the D0 data on the $W^\pm$-asymmetry in our analysis. 
To check this issue in a quantitative way we compare the D0 data on the $W^\pm$-asymmetry
with the predictions based on the results of present analysis. 
Despite a good description of the charged lepton data is achieved in our fit, the $W^\pm$-asymmetry data 
systematically undershoot the predictions, cf. Fig.~\ref{fig:d0w}. 
Therefore the $W^\pm$-asymmetry data would introduce a bias to our fit. 
A conceivable impact of those data on the PDFs can roughly be estimated from Fig.~\ref{fig:d0w}. 
It consists in a shift of the $d/u$ ratio at large $x$ by $1\sigma$ down and an
essential reduction of the error in the $d/u$ ratio, 
since the errors in the data at large pseudo-rapidity 
are significantly smaller the ones in predictions. 
Such a reduction of errors is also observed in the 
{\tt HERAFitter} analysis~\cite{Camarda:2015zba}.
This trend requires further clarification of the D0 data 
statistical significance on the information theory framework, since, as we have pointed out above, 
the same data sample is used to produce both the charged-lepton and the $W^\pm$-asymmetry distributions. 

Other recent PDF determinations are partially based on the data considered in our analysis,
cf. Table~\ref{tab:data-chi}, although a full coverage is not achieved 
anywhere else~\footnote{The Tevatron data on the charged $W^\pm$-asymmetry 
are used in the MMHT14 fit~\cite{Harland-Lang:2014zoa} instead.}. 
Despite this partial coverage of data in other PDF fits, the values of $\chi^2$ achieved in those fits
are generally bigger than ours. In particular, this concerns the 
ATLAS data of Ref.~\cite{Aad:2011dm} and the D0 electron data of 
Ref.~\cite{D0:2014kma}, which provide an essential constraints on the quark 
iso-spin asymmetry in our analysis. 
The value of $\chi^2$ obtained in the present analysis for the ATLAS data is also smaller than 
the one of the ABM12 fit~\cite{Alekhin:2013nda}, 
evidently due to the relaxed form of the small-$x$ iso-spin asymmetry employed.

\begin{figure}[tbh]
\centerline{
  \includegraphics[width=10cm]{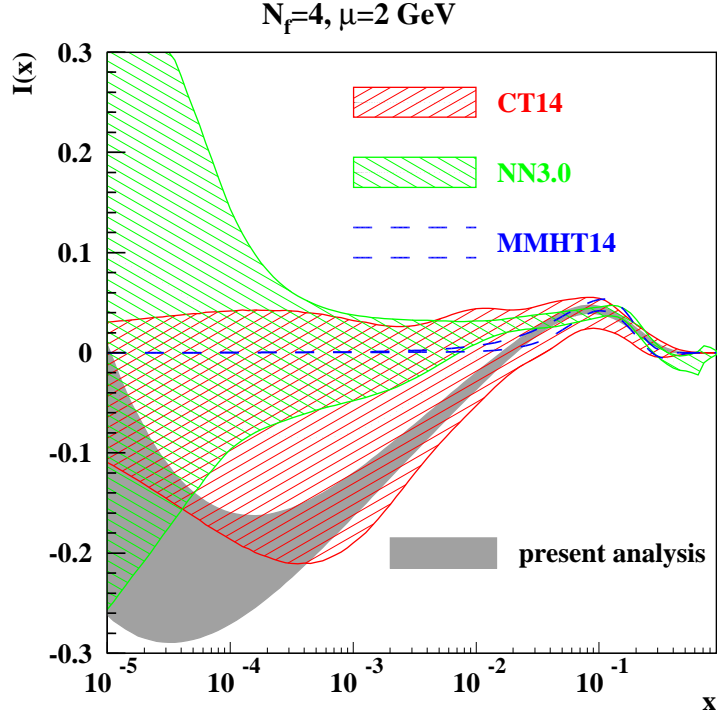}}
  \caption{\small  
  The $1\sigma$ band for 
  the iso-spin asymmetry of the sea $I(x)$ for the 4-flavor scheme
    at the scale of
    $\mu=2~{\rm GeV}$ as a function of the Bjorken $x$ obtained in the 
present fit (gray shaded area) 
in comparison with the corresponding ones obtained in the 
CT14~\cite{Dulat:2015mca} (red right-tilted hatch), 
MMHT14~\cite{Harland-Lang:2014zoa} (blue dashed lines), and
NN3.0~\cite{Ball:2014uwa} (green left-tilted hatch) analyses. 
    \label{fig:udmcomp}
}
\end{figure}

\begin{figure}[tbh]
\centerline{
  \includegraphics[width=10cm]{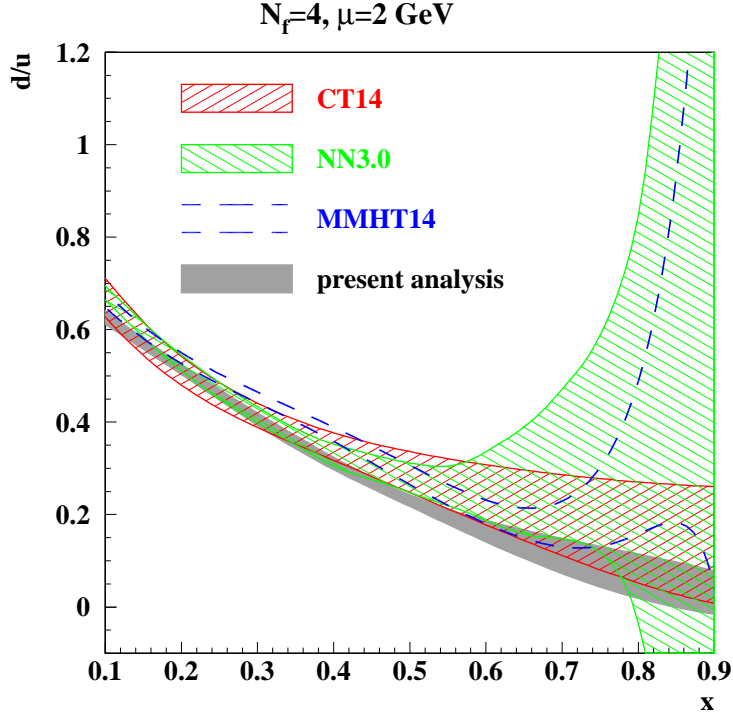}}
  \caption{\small Same as in Fig.~\ref{fig:udmcomp} for the ratio $d/u$.
    \label{fig:ducomp}
}
\end{figure}

A selected set of the NNLO quark distributions obtained 
in the present analysis is compared to the latest results of the
CT~\cite{Dulat:2015mca}, NNPDF~\cite{Ball:2014uwa}, and 
MMHT~\cite{Harland-Lang:2014zoa} PDF fits in Figs.~\ref{fig:udmcomp} 
and \ref{fig:ducomp}~\footnote{The {\tt HERAFitter} 
analysis~\cite{Camarda:2015zba} is performed to NLO accuracy in QCD only.
Therefore it is not considered in this comparison.}. 
In the MMHT14 analysis the sea iso-spin asymmetry $I(x)$ is parameterized 
in the Regge-like form Eq.~(\ref{eq:iso}), 
so it vanishes at small $x$. 
The CT14 and NN3.0 fits are based on a relaxed form of $I(x)$ at small $x$, 
and in our analysis. 
Due to the limited set of the $W^\pm$- and $Z$-boson collider data used in those fits, 
the CT14 and NN3.0 errors in $I(x)$
are significantly bigger than the ones obtained in our analysis.
However, the CT14 error band for $I(x)$ is comparable with ours, while the one
of NN3.0 goes higher at $x\sim 10^{-3}$. 
At $x\sim 0.1$ all determinations are consistent since this kinematics range is controlled by the fixed-target 
DY dimuon data, common for all four fits. 
The shape of $I(x)$ at large $x$ in CT14, MMHT14 and our analysis is controlled by the quark sum rule. 
Therefore it vanishes at $x=1$. 
A similar behavior is observed for the NN3.0 result, 
despite the author's declaration of a model-independent parametrization of PDFs. 
Evidently, some phase-space constraints are still applied in this case. 
However, a detailed clarification of this issue is impossible in view of the
fact that the explicit shape of the NN3.0 PDFs is unpublished. 
The ratio $d/u$ at large $x$ obtained in our analysis goes lower than other determinations, although it is 
consistent with them within uncertainties, which are much bigger than ours, 
especially for the MMHT14 and NN3.0 cases. 
As well as in the case of $I(x)$ this happens since a limited set of the
$W^\pm$- and $Z$-boson collider data is used in those fits. 
Besides, the cut of $W^2\gtrsim 13~{\rm GeV}^2$ on the invariant mass $W$ of the produced hadronic system 
imposed on the DIS data in the CT14, MMHT14, and NN3.0 analyses
removes the large-$x$ data for the deuteron target, which can provide an additional 
constraint on the $d/u$ ratio, as it happens in the ABM12 analysis~\cite{Alekhin:2013nda}. 
Since the $d/u$ ratio at large $x$ is basically uncontrolled in the MMHT14 and NN3.0 analyses, 
its central values differ from ours at $x=0.9$ by factor 
of $\sim 30$ and $\sim 20$, respectively. 
Such a spread obviously makes these PDFs inapplicable for a precision study of 
physics effects beyond Standard Model at large scales. 
Meanwhile, the MMHT14 and NN3.0 predictions are in clear disagreement with 
the forward $W^\pm$-boson production data by LHCb and D0, cf. Figs.~\ref{fig:tev} and \ref{fig:lhc}. 
This means, they can be consolidated 
with ours once additional $W^\pm$- and $Z$-boson collider data are included
into those analyses, in analogy to the recent CT14 results 
(see the difference between the CT10 and CT14 predictions in Figs.~\ref{fig:tev} and \ref{fig:lhc}).

The techniques used in the other PDF 
analyses~\cite{Harland-Lang:2014zoa,Ball:2014uwa,Dulat:2015mca}
to take into 
account the NNLO corrections to the DY differential distributions
are also different.
The MMHT14 analysis~\cite{Harland-Lang:2014zoa} is based on 
the NLO {\tt APPLGrid}~\cite{Carli:2010rw} output supplemented by NNLO 
$K$-factors computed with the codes {\tt FEWZ}~\cite{Li:2012wna} and 
{\tt DYNNLO}~\cite{Catani:2010en}.
For the most accurate recent data on the lepton asymmetry this approximation 
is insufficient since the NNLO corrections demonstrate a dependence on the 
lepton pseudo-rapidity with a spread beyond the data uncertainties,
cf. Fig.~\ref{fig:d0eacc}. In the NN3.0 analysis~\cite{Ball:2014uwa}  
the NNLO corrections are applied in the form of so-called 
$C$-factors~\cite{Ball:2011uy} including kinematic dependence of 
the NNLO corrections, however, still computed with the PDFs fixed. 
Finally, the CT14 analysis~\cite{Dulat:2015mca} is based on the 
{\tt ResBos} code~\cite{Balazs:1995nz}, which provides the QCD corrections
up to the NNLL accuracy only. In view of such a variety of tools 
a consolidation of the PDFs obtained in those fits also requires 
a thorough validation of the accuracy of the theoretical predictions used.

\begin{table}[th!]
\renewcommand{\arraystretch}{1.3}
\begin{center}                   
{\small                          
\begin{tabular}{|c|c|c|c|c|c|c|c|}   
\hline                           
\multicolumn{2}{|c|}{Experiment}                      
&ATLAS
&CMS  
&\multicolumn{2}{c|}{D0}
&\multicolumn{2}{c|}{LHCb}
\\
\hline                                                    
\multicolumn{2}{|c|}{$\sqrt s$~(TeV)}                      
&7                         
&7  
&\multicolumn{2}{c|}{1.96}
&{7}
&8                                                        
\\                                                        
\hline
\multicolumn{2}{|c|}{Final states} 
& $W^+\rightarrow l^+\nu$
& $W^+\rightarrow \mu^+\nu$
&$W^+\rightarrow \mu^+\nu$
& $W^+\rightarrow e^+\nu$
&$W^+\rightarrow \mu^+\nu$
& $Z\rightarrow e^+e^-$                                                        
\\
\multicolumn{2}{|c|}{ }                                          
& $W^-\rightarrow l^-\nu$
&$W^-\rightarrow \mu^-\nu$
&$W^-\rightarrow \mu^-\nu$
&$W^-\rightarrow e^-\nu$
&$W^-\rightarrow \mu^-\nu$
&                                                         
\\                                                        
\multicolumn{2}{|c|}{ }                                          
& $Z\rightarrow l^+l^-$
&                                                         
&
&                                                         
& $Z\rightarrow \mu^+\mu^-$
&                                                         
\\                                                        
\hline                           
\multicolumn{2}{|c|}{Reference}                      
&\cite{Aad:2011dm}                         
&\cite{Chatrchyan:2013mza}
&\cite{Abazov:2013rja}
&\cite{D0:2014kma}
&\cite{Aaij:2015gna}
&\cite{Aaij:2015vua}
\\                                                        
\hline                                                    
\multicolumn{2}{|c|}{$NDP$}
&30                      
&11  
&10
&13
&31
&17                            
\\                                                        
\hline
\multirow{2}{4em}{ }
& this work
 &29.8
 &$22.5$
 &16.9
 &18.0
&44.1
&18.2
\\
\cline{2-8}
&this work\footnote{The variants with all collider DY and $W^\pm$-boson data
  excluded except the one given.}
 &32.3
 &19.5(13.5\footnote{The value obtained assuming  
systematic uncertainties to be uncorrelated.} )
 &13.5
&9.5
 &34.7
&$19.1$
\\
\cline{2-8}
&ABM12~\cite{Alekhin:2013nda}
 &34.5
 &--
 &--
&--
 &--
&--
\\
\cline{2-8}
$\chi^2$ &CT14~\cite{Dulat:2015mca}
 &--\footnote{The ATLAS data on $W^\pm$- and $Z$-boson production cross sections are 
used in combination with the data on the lepton charge asymmetry. The value 
of $\chi^2/NDP=51/41=1.25$ is obtained for this sample.}
 &--~\footnote{Statistically less significant data with the cut of 
$P_T^{\mu}>35~{\rm GeV}$ are used.}
 &--
&34.7
 &--
&--
\\
\cline{2-8}
&{\tt HERAFitter}~\cite{Camarda:2015zba}
 &--
 &--
 &13
&19
 &--
&--
\\
\cline{2-8}
&MMHT14~\cite{Harland-Lang:2014zoa}
 &39
 &--
 &21
&--
 &--
&--
\\
\cline{2-8}
&NN3.0~\cite{Ball:2014uwa}
 &35.4
 &18.9
 &--
&--
 &--
&--
\\
\hline                                          
\end{tabular}
}
\caption{\label{tab:data-chi}
\small 
The values of $\chi^2$ obtained for the data samples considered in the present analysis, 
cf.~Table~\ref{tab:data-inp}, in comparison with the ones obtained in other PDF fits. 
The NNLO results are quoted for all cases, except for the {\tt HERAFitter} results~\cite{Camarda:2015zba}.} 
\end{center}
\end{table}

\section {Single-top hadronic production}

The results for the PDF fit have interesting implications in case of theory
predictions for the production of single top-quarks at the LHC.
The $t$-channel single (anti)top-quark cross sections 
$\sigma_t$ and $\sigma_{\bar t}$ have been measured at the LHC for $\sqrt{s} = 7$~TeV by ATLAS~\cite{Aad:2014fwa}
and for $\sqrt{s} = 8$~TeV by CMS~\cite{Khachatryan:2014iya}.
The inclusive cross sections at NLO in QCD can be conveniently computed 
with the {\tt Hathor} library~\cite{Aliev:2010zk,Kant:2014oha}, which features 
$t$-channel, $s$-channel and $Wt$- single top-quark production and 
employs the hard partonic cross sections at NLO in QCD based on Refs.~\cite{Harris:2002md,Campbell:2004ch} 
for the $t$-channel process.

For the PDFs of the present analysis as well as the sets discussed in the
previous Section, i.e., ABM12, CT10, CT14, MMHT, NN3.0, 
we evaluate the cross sections $\sigma_t$ and $\sigma_{\bar t}$ 
for the $t$-channel production of single (anti)top-quarks along with 
their ratio $R_t=\sigma_t/\sigma_{\bar t}$.
The cross sections are computed for running top-quark masses in the \msbar
scheme according to the algorithm described in Ref.~\cite{Langenfeld:2009wd}.
We choose $m_t(m_t) = 163$ GeV in the \msbar scheme, which
corresponds to a pole mass $m_t^{\rm pole} = 172.48$ GeV using the conversion
to the on-shell scheme at three loops, cf. Ref.~\cite{Marquard:2015qpa}.

\begin{figure}[tbh]
  \begin{center}
  \includegraphics[width=7.75cm]{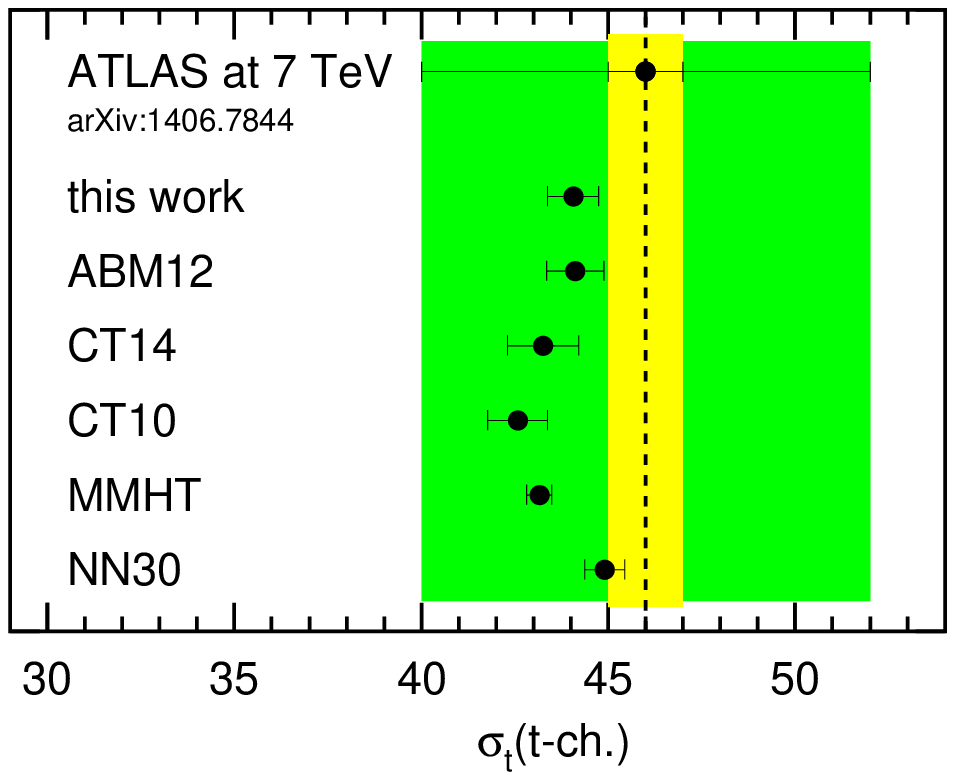}
  \includegraphics[width=7.75cm]{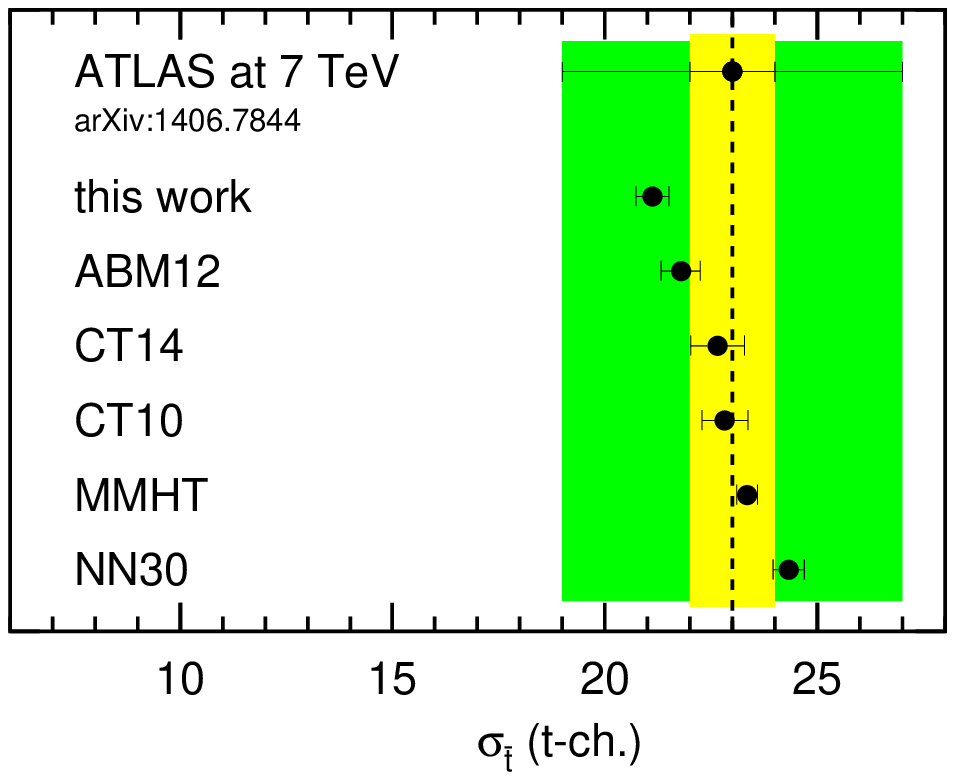}
  \end{center}
  \vspace*{-5mm}
  \caption{\small
    \label{fig:sigt7}
    Cross sections together with their $1\sigma$ PDF uncertainties
    for the $t$-channel production of single (anti)top-quarks in $pp$ collision  
    at $\sqrt s = 7$~TeV in comparison to ATLAS data~\cite{Aad:2014fwa}
    for a \msbar mass $m_t(m_t) = 163$~GeV at the scale
    $\mu_R=\mu_F=m_t(m_t)$ with PDF sets are taken at NNLO.
    The inner (yellow) band denotes the statistical uncertainty 
    and the outer (green) band the combined uncertainty due to statistics and systematics.
  }
\end{figure}

\begin{figure}[tbh]
  \begin{center}
  \includegraphics[width=7.75cm]{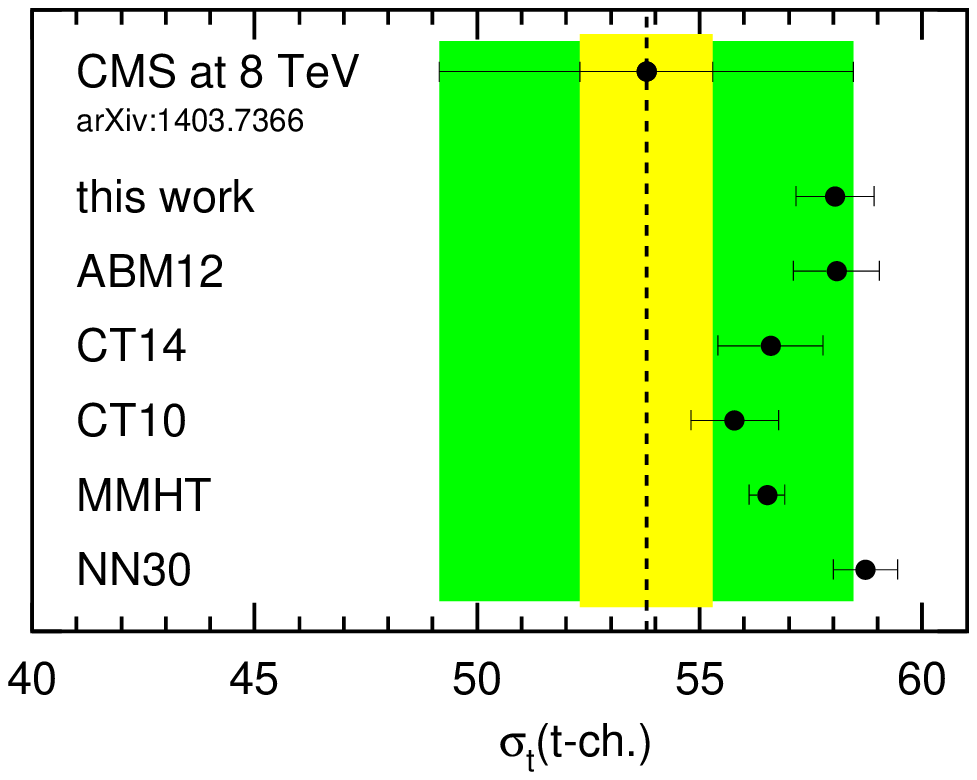}
  \includegraphics[width=7.75cm]{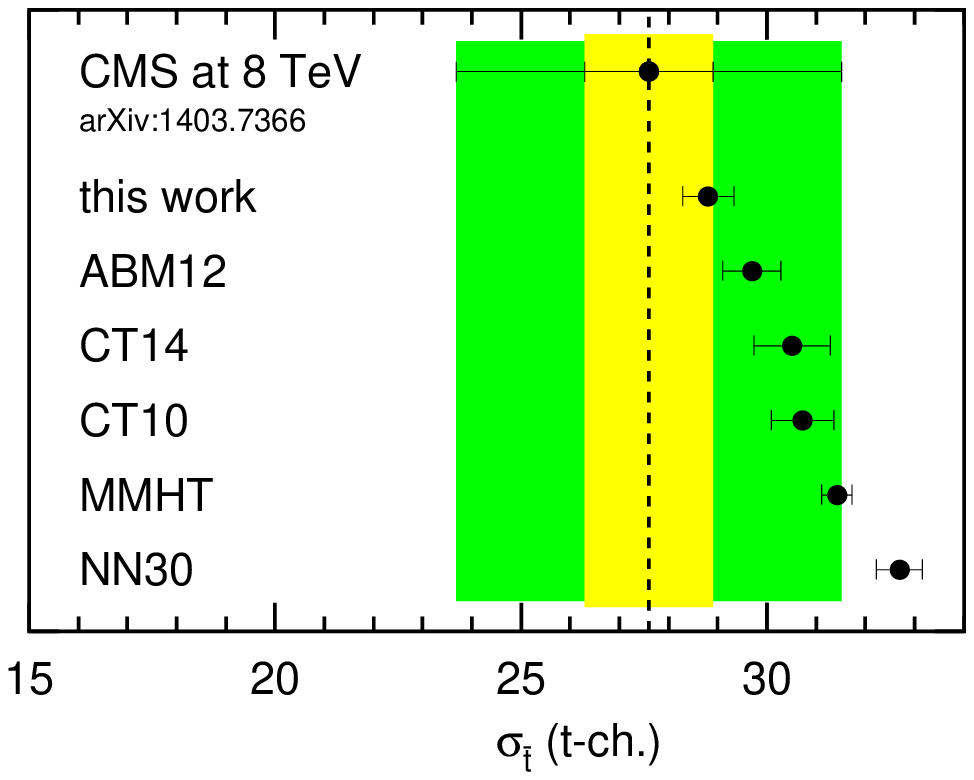}
  \end{center}
  \vspace*{-5mm}
  \caption{\small
    \label{fig:sigt8}
    Same as Fig.~\ref{fig:sigt7} for $pp$ collision  
    at $\sqrt s = 8$~TeV in comparison to CMS data~\cite{Khachatryan:2014iya}.
  }
\end{figure}

In Figs.~\ref{fig:sigt7} and \ref{fig:sigt8} we display 
the cross sections $\sigma_t$ and $\sigma_{\bar t}$ 
at $\sqrt s = 7$~TeV and $\sqrt s = 8$~TeV, respectively,  
using $m_t(m_t) = 163$ GeV at the nominal values for the scales, $\mu_R=\mu_F=m_t(m_t)$.
The central value for each PDF set is complemented by the symmetrized PDF and, if
applicable, $\alpha_s$ uncertainties.
It is obvious from Figs.~\ref{fig:sigt7} and \ref{fig:sigt8} that 
within the current experimental uncertainties all predictions agree with 
data from ATLAS~\cite{Aad:2014fwa} and CMS~\cite{Khachatryan:2014iya}.
Moreover, the results demonstrate that the cross section predictions are quite
stable against higher order perturbative corrections.
Predictions based on PDF sets taken either at NLO or at NNLO agree with uncertainties.
Also, the known NNLO QCD corrections~\cite{Brucherseifer:2014ama} to the hard partonic cross section are quite small. 
Interestingly, in the case with no cut on the transverse momentum of the top
quark they are negative, 
with values reported~\cite{Brucherseifer:2014ama} 
for the ratios $\sigma_t^{\rm NNLO}/\sigma_t^{\rm NLO} \simeq -1.6\%$ 
and $\sigma_{\bar t}^{\rm NNLO}/\sigma_{\bar t}^{\rm NLO} \simeq -1.3\%$ for
$pp$-collisions at $\sqrt s = 8$~TeV for a pole mass $m_t^{\rm pole} = 173.2$ GeV.
Note, that the mass dependence of the single top-quark cross sections is very
mild, cf. Ref.~\cite{Kant:2014oha}, so that these conclusions are directly
applicable to our analysis as well.

\begin{figure}[tbh]
  \begin{center}
  \includegraphics[width=7.70cm]{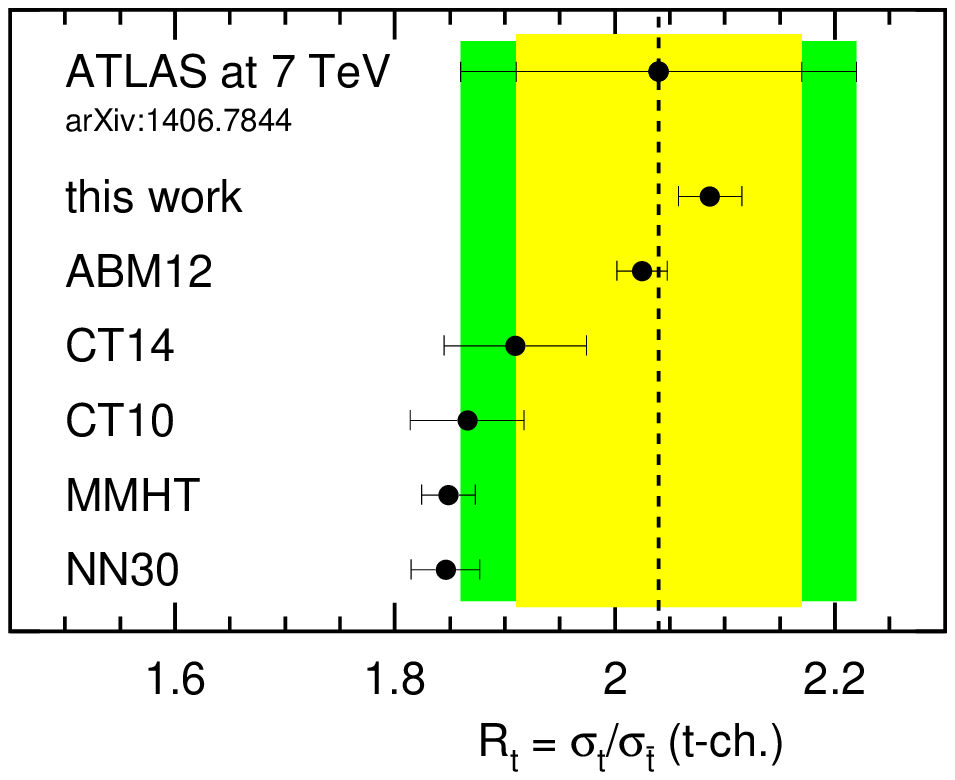}
  \includegraphics[width=7.95cm]{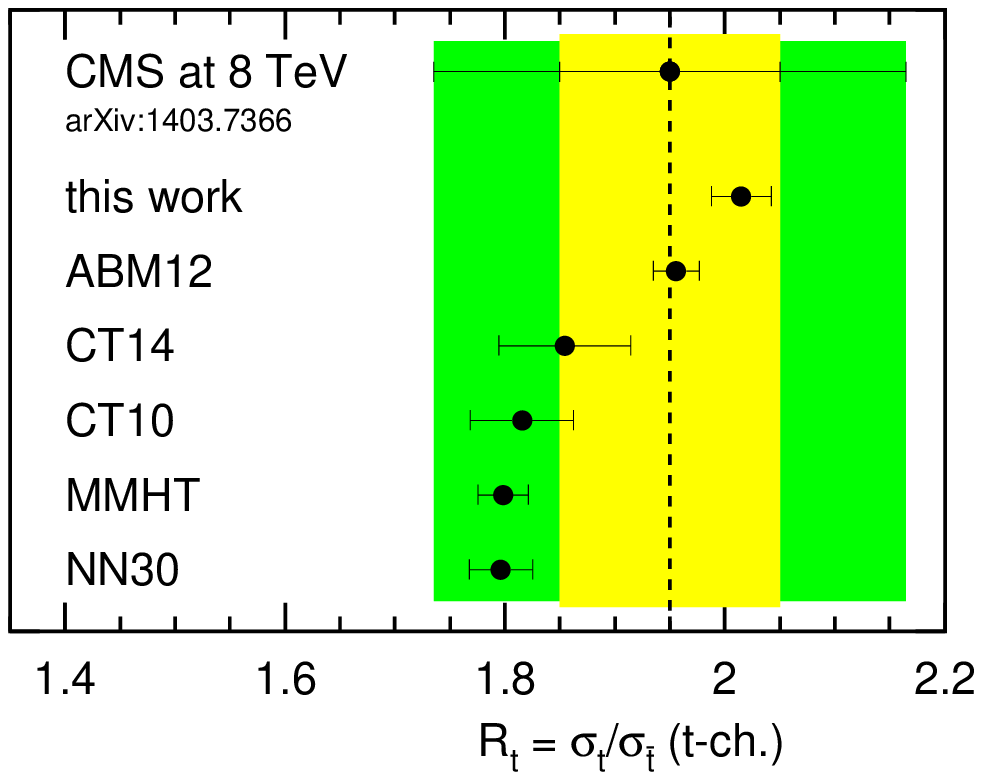}
  \end{center}
  \vspace*{-5mm}
  \caption{\small
    \label{fig:ratio7+8}
    Same as in Fig.~\ref{fig:sigt7} for the 
    ratio of cross sections $R_t=\sigma_t/\sigma_{\bar t}$ 
    in comparison to ATLAS data~\cite{Aad:2014fwa} at $\sqrt s = 7$~TeV (left) 
    and  to CMS data~\cite{Khachatryan:2014iya} at $\sqrt s = 8$~TeV (right).
  }
\end{figure}

In Fig.~\ref{fig:ratio7+8} we show results for the ratio $R_t=\sigma_t/\sigma_{\bar t}$ of the cross sections.
In this quantity, theory uncertainties basically cancel, making $R_t$ a very 
sensitive probe for the $d/u$ ratio in PDFs at large $x$.
From the ATLAS and CMS data plotted in Fig.~\ref{fig:ratio7+8} is also evident, 
that uncertainties due to experimental systematics have almost completely
canceled, so that current measurements of $R_t$ are limited by statistics.
All predictions for $R_t$ with the various PDF sets agree with the 
ATLAS~\cite{Aad:2014fwa} and CMS~\cite{Khachatryan:2014iya} data 
within uncertainties, although the results for CT10, CT14, MMHT and NN3.0 are
shifted systematically to the edge of the $1\sigma$ interval of the 
experimental uncertainty band. 
The numbers for ABM12 and this work do agree very well with the central values for $R_t$ 
measured by ATLAS and CMS.
With much higher rates and, as a consequence, improved statistics for top-quark production in the high energy run 
at $\sqrt s = 13$~TeV, the prospects for single top-quark production to serve
as a standard candle process are very good.

\section{Conclusions}

We have used the currently available data for 
hadro-production of $W^\pm$- and $Z$-bosons at the LHC and Tevatron
from the experiments ATLAS, CMS, D0 and LHCb 
to constrain the light quark flavor distributions in the proton.
Due to the data's kinematic range extending to large rapidities important new 
information on the light quark PDFs over a wide range in $x$ has been obtained.
Given the unprecedented experimental precision with total systematic errors often well below 1\% 
theory comparisons at NNLO accuracy in QCD are mandatory.
The analysis is based on fully differential QCD predictions at NNLO for the 
rapidity distributions of the leptons from the $W^\pm$- and $Z$-boson decay.
The PDF fit requires fast and reliable theory predictions at per mil level numerical integration accuracy 
which poses a challenge for currently available codes. 
In the present analysis the choice has been the publicly available code {\tt FEWZ} at the expense of 
typical CPU times necessary to accomplish these computations of $O(10^{4}{\rm h})$ for each data set.

The new data considered in the present analysis allow for a refinement 
of the parameterization for the $u$- and $d$-quark distributions.
The PDF fit has been performed as a variant of the ABM12 analysis, 
which is based on the world DIS data, on DY dimuon data from fixed targets 
and on early LHC data for hadro-production of $W^\pm$- and $Z$-bosons.
Accommodation of the new data in the fit can be achieved by exploiting 
the parametric freedom either in $d$-quark distribution at large $x$ or in the
$\bar d$-quark distribution at small $x$. 
The agreement with data for forward $W^\pm$-boson production 
at large rapidities $\eta \sim 2.5 \div 3$ for D0 
and in the range $\eta \sim 3 \div 4.5$ for LHCb 
has suggested a modification of the $\bar d$-quark distribution at small $x$.
By relaxing the commonly chosen parametric ansatz for the iso-spin asymmetric sea 
$x(\bar d - \bar u)$ at small $x$, both D0 and LHCb data can be consistently accommodated in the fit.
This leads to a non-vanishing $x(\bar d - \bar u)$ asymmetry at values 
$x\sim 10^{-4}$ and a delayed onset of the generally anticipated Regge asymptotics at small $x$, 
a feature which is not entirely without precedence in perturbative QCD.

In the analysis great care has been taken to consider only data sets which are
mutually consistent and we have demonstrated that there are issues with the LHCb data.
In checking the consistency of LHCb data for $Z$-boson production, we have noted
tensions among the various data sets, which has lead to the exclusion of the 
$Z\to e^+e^-$ LHCb data taken at $\sqrt{s} = 7$~TeV from the fit.
We have also observed that different data for the decay of forward $W^\pm$-bosons 
published by LHCb are not consistent with each other 
regarding the shape and the treatment of final state radiation to account for QED corrections. 
This has lead us to reconsider the quoted uncertainties for this source.
Finally, the LHCb data for $W^\pm$-boson production display 
some tension with the respective D0 data sets, which is noticeable from a 
deterioration of the $\chi^2$ values in the combined fit.
In that case, however, current information is insufficient to resolve the
observed discrepancies which show up as fluctuations at the level of 1$\sigma$
in favor either of the D0 or the LHCb experiment.

The results of the present analysis and the data have been compared 
to predictions obtained with other global PDF sets.
The theory predictions at NNLO in QCD for the charged lepton asymmetry from $W^\pm$-decays 
based on CT10, MMHT or the so-called unbiased fit NN3.0
undershoot the data at large rapidities dramatically,  
leading to poor compatibility of those predictions with D0 and LHCb data in the forward region.
This indicates a significant constraining power of the data for those global
fits, as has been pointed out in a recent {\tt HERAFitter} analysis as well.

As an application to collider phenomenology, 
the fit results have been used to predict the cross sections 
for the production of single (anti)top-quarks at the LHC, $\sigma_t$, $\sigma_{\bar t}$ 
and their ratio $R_t=\sigma_t/\sigma_{\bar t}$, 
demonstrating very good agreement with existing data and underpinning the
potential of this process to serve as a standard candle.

In summary, the analysis shows the gain in knowledge on the light flavor PDFs from current LHC and Tevatron data.
We expect further improvements from data on hadro-production 
of $W^\pm$- and $Z$-bosons as well as from high statistics measurements of single (anti)top-quark production
in run II of the LHC at $\sqrt{s} = 13$~TeV.
Additional advances in the small-$x$ regime can be expected from the proposed Large Hadron Electron Collider (LHeC)~\cite{AbelleiraFernandez:2012cc}, 
which is to collide an electron beam up to possibly 140 GeV with the intense
hadron beams of the LHC, including deuterons, 
and which extends the currently covered kinematic range in Bjorken-$x$ significantly.

\subsection*{Acknowledgments}
We are grateful to Stefano Camarda for clarification of the D0 data details, 
to Uta Klein for useful discussions of {\tt FEWZ} features, 
and to Ronan J. Wallace for communication on LHCb data details,  
This work has been supported by Bundesministerium f\"ur Bildung und Forschung through contract (05H15GUCC1).

{\footnotesize                                                          
%\bibliography{abib1}

\begin{thebibliography}{40}
\expandafter\ifx\csname natexlab\endcsname\relax\def\natexlab#1{#1}\fi
\expandafter\ifx\csname bibnamefont\endcsname\relax
  \def\bibnamefont#1{#1}\fi
\expandafter\ifx\csname bibfnamefont\endcsname\relax
  \def\bibfnamefont#1{#1}\fi
\expandafter\ifx\csname citenamefont\endcsname\relax
  \def\citenamefont#1{#1}\fi
\expandafter\ifx\csname url\endcsname\relax
  \def\url#1{\texttt{#1}}\fi
\expandafter\ifx\csname urlprefix\endcsname\relax\def\urlprefix{URL }\fi
\providecommand{\bibinfo}[2]{#2}
\providecommand{\eprint}[2][]{\url{#2}}

\bibitem[{\citenamefont{Chatrchyan et~al.}(2014)}]{Chatrchyan:2013mza}
\bibinfo{author}{\bibfnamefont{S.}~\bibnamefont{Chatrchyan}}
  \bibnamefont{et~al.} (\bibinfo{collaboration}{CMS}),
  \bibinfo{journal}{Phys.Rev.} \textbf{\bibinfo{volume}{D90}},
  \bibinfo{pages}{032004} (\bibinfo{year}{2014}), \eprint{1312.6283}.

\bibitem[{\citenamefont{Aaij et~al.}(2014)}]{Aaij:2014wba}
\bibinfo{author}{\bibfnamefont{R.}~\bibnamefont{Aaij}} \bibnamefont{et~al.}
  (\bibinfo{collaboration}{LHCb}), \bibinfo{journal}{JHEP}
  \textbf{\bibinfo{volume}{1412}}, \bibinfo{pages}{079} (\bibinfo{year}{2014}),
  \eprint{1408.4354}.

\bibitem[{\citenamefont{Abazov et~al.}(2013)}]{Abazov:2013rja}
\bibinfo{author}{\bibfnamefont{V.~M.} \bibnamefont{Abazov}}
  \bibnamefont{et~al.} (\bibinfo{collaboration}{D0}),
  \bibinfo{journal}{Phys.Rev.} \textbf{\bibinfo{volume}{D88}},
  \bibinfo{pages}{091102} (\bibinfo{year}{2013}), \eprint{1309.2591}.

\bibitem[{\citenamefont{Abazov et~al.}(2015)}]{D0:2014kma}
\bibinfo{author}{\bibfnamefont{V.~M.} \bibnamefont{Abazov}}
  \bibnamefont{et~al.} (\bibinfo{collaboration}{D0}),
  \bibinfo{journal}{Phys.Rev.} \textbf{\bibinfo{volume}{D91}},
  \bibinfo{pages}{032007} (\bibinfo{year}{2015}), \eprint{1412.2862}.

\bibitem[{\citenamefont{Alekhin et~al.}(2015)\citenamefont{Alekhin,
  Bl{\"u}mlein, Caminada, Lipka, Lohwasser et~al.}}]{Alekhin:2014sya}
\bibinfo{author}{\bibfnamefont{S.}~\bibnamefont{Alekhin}},
  \bibinfo{author}{\bibfnamefont{J.}~\bibnamefont{Bl{\"u}mlein}},
  \bibinfo{author}{\bibfnamefont{L.}~\bibnamefont{Caminada}},
  \bibinfo{author}{\bibfnamefont{K.}~\bibnamefont{Lipka}},
  \bibinfo{author}{\bibfnamefont{K.}~\bibnamefont{Lohwasser}},
  \bibnamefont{et~al.}, \bibinfo{journal}{Phys.Rev.}
  \textbf{\bibinfo{volume}{D91}}, \bibinfo{pages}{094002}
  (\bibinfo{year}{2015}), \eprint{1404.6469}.

\bibitem[{\citenamefont{Alekhin et~al.}(2010)\citenamefont{Alekhin,
  Bl{\"u}mlein, Klein, and Moch}}]{Alekhin:2009ni}
\bibinfo{author}{\bibfnamefont{S.}~\bibnamefont{Alekhin}},
  \bibinfo{author}{\bibfnamefont{J.}~\bibnamefont{Bl{\"u}mlein}},
  \bibinfo{author}{\bibfnamefont{S.}~\bibnamefont{Klein}}, \bibnamefont{and}
  \bibinfo{author}{\bibfnamefont{S.}~\bibnamefont{Moch}},
  \bibinfo{journal}{Phys.Rev.} \textbf{\bibinfo{volume}{D81}},
  \bibinfo{pages}{014032} (\bibinfo{year}{2010}), \eprint{0908.2766}.

\bibitem[{\citenamefont{Alekhin et~al.}(2012)\citenamefont{Alekhin,
  Bl{\"u}mlein, and Moch}}]{Alekhin:2012ig}
\bibinfo{author}{\bibfnamefont{S.}~\bibnamefont{Alekhin}},
  \bibinfo{author}{\bibfnamefont{J.}~\bibnamefont{Bl{\"u}mlein}},
  \bibnamefont{and} \bibinfo{author}{\bibfnamefont{S.}~\bibnamefont{Moch}},
  \bibinfo{journal}{Phys.Rev.} \textbf{\bibinfo{volume}{D86}},
  \bibinfo{pages}{054009} (\bibinfo{year}{2012}), \eprint{1202.2281}.

\bibitem[{\citenamefont{Alekhin et~al.}(2014)\citenamefont{Alekhin,
  Bl{\"u}mlein, and Moch}}]{Alekhin:2013nda}
\bibinfo{author}{\bibfnamefont{S.}~\bibnamefont{Alekhin}},
  \bibinfo{author}{\bibfnamefont{J.}~\bibnamefont{Bl{\"u}mlein}},
  \bibnamefont{and} \bibinfo{author}{\bibfnamefont{S.}~\bibnamefont{Moch}},
  \bibinfo{journal}{Phys.Rev.} \textbf{\bibinfo{volume}{D89}},
  \bibinfo{pages}{054028} (\bibinfo{year}{2014}), \eprint{1310.3059}.

\bibitem[{\citenamefont{Gao et~al.}(2014)\citenamefont{Gao, Guzzi, Huston, Lai,
  Li et~al.}}]{Gao:2013xoa}
\bibinfo{author}{\bibfnamefont{J.}~\bibnamefont{Gao}},
  \bibinfo{author}{\bibfnamefont{M.}~\bibnamefont{Guzzi}},
  \bibinfo{author}{\bibfnamefont{J.}~\bibnamefont{Huston}},
  \bibinfo{author}{\bibfnamefont{H.-L.} \bibnamefont{Lai}},
  \bibinfo{author}{\bibfnamefont{Z.}~\bibnamefont{Li}}, \bibnamefont{et~al.},
  \bibinfo{journal}{Phys.Rev.} \textbf{\bibinfo{volume}{D89}},
  \bibinfo{pages}{033009} (\bibinfo{year}{2014}), \eprint{1302.6246}.

\bibitem[{\citenamefont{Dulat et~al.}(2015)\citenamefont{Dulat, Hou, Gao,
  Guzzi, Huston, Nadolsky, Pumplin, Schmidt, Stump, and Yuan}}]{Dulat:2015mca}
\bibinfo{author}{\bibfnamefont{S.}~\bibnamefont{Dulat}},
  \bibinfo{author}{\bibfnamefont{T.~J.} \bibnamefont{Hou}},
  \bibinfo{author}{\bibfnamefont{J.}~\bibnamefont{Gao}},
  \bibinfo{author}{\bibfnamefont{M.}~\bibnamefont{Guzzi}},
  \bibinfo{author}{\bibfnamefont{J.}~\bibnamefont{Huston}},
  \bibinfo{author}{\bibfnamefont{P.}~\bibnamefont{Nadolsky}},
  \bibinfo{author}{\bibfnamefont{J.}~\bibnamefont{Pumplin}},
  \bibinfo{author}{\bibfnamefont{C.}~\bibnamefont{Schmidt}},
  \bibinfo{author}{\bibfnamefont{D.}~\bibnamefont{Stump}}, \bibnamefont{and}
  \bibinfo{author}{\bibfnamefont{C.~P.} \bibnamefont{Yuan}}
  (\bibinfo{year}{2015}), \eprint{1506.07443}.

\bibitem[{\citenamefont{Harland-Lang et~al.}(2015)\citenamefont{Harland-Lang,
  Martin, Motylinski, and Thorne}}]{Harland-Lang:2014zoa}
\bibinfo{author}{\bibfnamefont{L.}~\bibnamefont{Harland-Lang}},
  \bibinfo{author}{\bibfnamefont{A.}~\bibnamefont{Martin}},
  \bibinfo{author}{\bibfnamefont{P.}~\bibnamefont{Motylinski}},
  \bibnamefont{and} \bibinfo{author}{\bibfnamefont{R.}~\bibnamefont{Thorne}},
  \bibinfo{journal}{Eur.Phys.J.} \textbf{\bibinfo{volume}{C75}},
  \bibinfo{pages}{204} (\bibinfo{year}{2015}), \eprint{1412.3989}.

\bibitem[{\citenamefont{Ball et~al.}(2015)}]{Ball:2014uwa}
\bibinfo{author}{\bibfnamefont{R.~D.} \bibnamefont{Ball}} \bibnamefont{et~al.}
  (\bibinfo{collaboration}{NNPDF}), \bibinfo{journal}{JHEP}
  \textbf{\bibinfo{volume}{1504}}, \bibinfo{pages}{040} (\bibinfo{year}{2015}),
  \eprint{1410.8849}.

\bibitem[{\citenamefont{Camarda et~al.}(2015)}]{Camarda:2015zba}
\bibinfo{author}{\bibfnamefont{S.}~\bibnamefont{Camarda}} \bibnamefont{et~al.}
  (\bibinfo{collaboration}{HERAFitter developers' Team})
  (\bibinfo{year}{2015}), \eprint{1503.05221}.

\bibitem[{\citenamefont{Brucherseifer et~al.}(2014)\citenamefont{Brucherseifer,
  Caola, and Melnikov}}]{Brucherseifer:2014ama}
\bibinfo{author}{\bibfnamefont{M.}~\bibnamefont{Brucherseifer}},
  \bibinfo{author}{\bibfnamefont{F.}~\bibnamefont{Caola}}, \bibnamefont{and}
  \bibinfo{author}{\bibfnamefont{K.}~\bibnamefont{Melnikov}},
  \bibinfo{journal}{Phys.Lett.} \textbf{\bibinfo{volume}{B736}},
  \bibinfo{pages}{58} (\bibinfo{year}{2014}), \eprint{1404.7116}.

\bibitem[{\citenamefont{Aad et~al.}(2014)}]{Aad:2014fwa}
\bibinfo{author}{\bibfnamefont{G.}~\bibnamefont{Aad}} \bibnamefont{et~al.}
  (\bibinfo{collaboration}{ATLAS}), \bibinfo{journal}{Phys.Rev.}
  \textbf{\bibinfo{volume}{D90}}, \bibinfo{pages}{112006}
  (\bibinfo{year}{2014}), \eprint{1406.7844}.

\bibitem[{\citenamefont{Khachatryan et~al.}(2014)}]{Khachatryan:2014iya}
\bibinfo{author}{\bibfnamefont{V.}~\bibnamefont{Khachatryan}}
  \bibnamefont{et~al.} (\bibinfo{collaboration}{CMS}), \bibinfo{journal}{JHEP}
  \textbf{\bibinfo{volume}{1406}}, \bibinfo{pages}{090} (\bibinfo{year}{2014}),
  \eprint{1403.7366}.

\bibitem{Alekhin:2010dd}
   S.~Alekhin, J.~Blumlein, P.~Jimenez-Delgado, S.~Moch and E.~Reya,
   %``NNLO Benchmarks for Gauge and Higgs Boson Production at TeV Hadron 
Colliders,''
   Phys.\ Lett.\ B {\bf 697} (2011) 127
   [arXiv:1011.6259 [hep-ph]].
   %%CITATION = ARXIV:1011.6259;%%
   %47 citations counted in INSPIRE as of 31 Aug 2015

\bibitem[{\citenamefont{Aaij et~al.}(2015{\natexlab{a}})}]{Aaij:2015gna}
\bibinfo{author}{\bibfnamefont{R.}~\bibnamefont{Aaij}} \bibnamefont{et~al.}
  (\bibinfo{collaboration}{LHCb}) (\bibinfo{year}{2015}{\natexlab{a}}),
  \eprint{1505.07024}.

\bibitem[{\citenamefont{Aaij et~al.}(2015{\natexlab{b}})}]{Aaij:2015vua}
\bibinfo{author}{\bibfnamefont{R.}~\bibnamefont{Aaij}} \bibnamefont{et~al.}
  (\bibinfo{collaboration}{LHCb}), \bibinfo{journal}{JHEP}
  \textbf{\bibinfo{volume}{1505}}, \bibinfo{pages}{109}
  (\bibinfo{year}{2015}{\natexlab{b}}), \eprint{1503.00963}.

\bibitem[{\citenamefont{Chatrchyan et~al.}(2012)}]{Chatrchyan:2012xt}
\bibinfo{author}{\bibfnamefont{S.}~\bibnamefont{Chatrchyan}}
  \bibnamefont{et~al.} (\bibinfo{collaboration}{CMS}),
  \bibinfo{journal}{Phys.Rev.Lett.} \textbf{\bibinfo{volume}{109}},
  \bibinfo{pages}{111806} (\bibinfo{year}{2012}), \eprint{1206.2598}.

\bibitem[{\citenamefont{Aaij et~al.}(2012)}]{Aaij:2012vn}
\bibinfo{author}{\bibfnamefont{R.}~\bibnamefont{Aaij}} \bibnamefont{et~al.}
  (\bibinfo{collaboration}{LHCb}), \bibinfo{journal}{JHEP}
  \textbf{\bibinfo{volume}{1206}}, \bibinfo{pages}{058} (\bibinfo{year}{2012}),
  \eprint{1204.1620}.

\bibitem[{\citenamefont{Aad et~al.}(2012)}]{Aad:2011dm}
\bibinfo{author}{\bibfnamefont{G.}~\bibnamefont{Aad}} \bibnamefont{et~al.}
  (\bibinfo{collaboration}{ATLAS}), \bibinfo{journal}{Phys.Rev.}
  \textbf{\bibinfo{volume}{D85}}, \bibinfo{pages}{072004}
  (\bibinfo{year}{2012}), \eprint{1109.5141}.

\bibitem[{\citenamefont{Aaij et~al.}(2013)}]{Aaij:2012mda}
\bibinfo{author}{\bibfnamefont{R.}~\bibnamefont{Aaij}} \bibnamefont{et~al.}
  (\bibinfo{collaboration}{LHCb}), \bibinfo{journal}{JHEP}
  \textbf{\bibinfo{volume}{1302}}, \bibinfo{pages}{106} (\bibinfo{year}{2013}),
  \eprint{1212.4620}.

\bibitem[{\citenamefont{Li and Petriello}(2012)}]{Li:2012wna}
\bibinfo{author}{\bibfnamefont{Y.}~\bibnamefont{Li}} \bibnamefont{and}
  \bibinfo{author}{\bibfnamefont{F.}~\bibnamefont{Petriello}},
  \bibinfo{journal}{Phys.Rev.} \textbf{\bibinfo{volume}{D86}},
  \bibinfo{pages}{094034} (\bibinfo{year}{2012}), \eprint{1208.5967}.

\bibitem[{\citenamefont{Gavin et~al.}(2013)\citenamefont{Gavin, Li, Petriello,
  and Quackenbush}}]{Gavin:2012sy}
\bibinfo{author}{\bibfnamefont{R.}~\bibnamefont{Gavin}},
  \bibinfo{author}{\bibfnamefont{Y.}~\bibnamefont{Li}},
  \bibinfo{author}{\bibfnamefont{F.}~\bibnamefont{Petriello}},
  \bibnamefont{and}
  \bibinfo{author}{\bibfnamefont{S.}~\bibnamefont{Quackenbush}},
  \bibinfo{journal}{Comput.Phys.Commun.} \textbf{\bibinfo{volume}{184}},
  \bibinfo{pages}{208} (\bibinfo{year}{2013}), \eprint{1201.5896}.

\bibitem[{\citenamefont{Golonka and Was}(2006)}]{Golonka:2005pn}
\bibinfo{author}{\bibfnamefont{P.}~\bibnamefont{Golonka}} \bibnamefont{and}
  \bibinfo{author}{\bibfnamefont{Z.}~\bibnamefont{Was}},
  \bibinfo{journal}{Eur.Phys.J.} \textbf{\bibinfo{volume}{C45}},
  \bibinfo{pages}{97} (\bibinfo{year}{2006}), \eprint{hep-ph/0506026}.

\bibitem[{\citenamefont{Bahr et~al.}(2008)\citenamefont{Bahr, Gieseke, Gigg,
  Grellscheid, Hamilton et~al.}}]{Bahr:2008pv}
\bibinfo{author}{\bibfnamefont{M.}~\bibnamefont{Bahr}},
  \bibinfo{author}{\bibfnamefont{S.}~\bibnamefont{Gieseke}},
  \bibinfo{author}{\bibfnamefont{M.}~\bibnamefont{Gigg}},
  \bibinfo{author}{\bibfnamefont{D.}~\bibnamefont{Grellscheid}},
  \bibinfo{author}{\bibfnamefont{K.}~\bibnamefont{Hamilton}},
  \bibnamefont{et~al.}, \bibinfo{journal}{Eur.Phys.J.}
  \textbf{\bibinfo{volume}{C58}}, \bibinfo{pages}{639} (\bibinfo{year}{2008}),
  \eprint{0803.0883}.

\bibitem[{\citenamefont{Sjostrand et~al.}(2008)\citenamefont{Sjostrand, Mrenna,
  and Skands}}]{Sjostrand:2007gs}
\bibinfo{author}{\bibfnamefont{T.}~\bibnamefont{Sjostrand}},
  \bibinfo{author}{\bibfnamefont{S.}~\bibnamefont{Mrenna}}, \bibnamefont{and}
  \bibinfo{author}{\bibfnamefont{P.~Z.} \bibnamefont{Skands}},
  \bibinfo{journal}{Comput.Phys.Commun.} \textbf{\bibinfo{volume}{178}},
  \bibinfo{pages}{852} (\bibinfo{year}{2008}), \eprint{0710.3820}.

\bibitem[{\citenamefont{Kluge et~al.}(2006)\citenamefont{Kluge, Rabbertz, and
  Wobisch}}]{Kluge:2006xs}
\bibinfo{author}{\bibfnamefont{T.}~\bibnamefont{Kluge}},
  \bibinfo{author}{\bibfnamefont{K.}~\bibnamefont{Rabbertz}}, \bibnamefont{and}
  \bibinfo{author}{\bibfnamefont{M.}~\bibnamefont{Wobisch}}, in
  \emph{\bibinfo{booktitle}{{Deep inelastic scattering. Proceedings, 14th
  International Workshop, DIS 2006, Tsukuba, Japan, April 20-24, 2006}}}
  (\bibinfo{year}{2006}), pp. \bibinfo{pages}{483--486},
  \eprint{hep-ph/0609285},
  \urlprefix\url{http://lss.fnal.gov/cgi-bin/find_paper.pl?conf-06-352}.

\bibitem[{\citenamefont{Kulagin and Petti}(2006)}]{Kulagin:2004ie}
\bibinfo{author}{\bibfnamefont{S.~A.} \bibnamefont{Kulagin}} \bibnamefont{and}
  \bibinfo{author}{\bibfnamefont{R.}~\bibnamefont{Petti}},
  \bibinfo{journal}{Nucl.Phys.} \textbf{\bibinfo{volume}{A765}},
  \bibinfo{pages}{126} (\bibinfo{year}{2006}), \eprint{hep-ph/0412425}.

\bibitem[{\citenamefont{Abazov et~al.}(2014)}]{Abazov:2013dsa}
\bibinfo{author}{\bibfnamefont{V.~M.} \bibnamefont{Abazov}}
  \bibnamefont{et~al.} (\bibinfo{collaboration}{D0}), \bibinfo{journal}{Phys.
  Rev. Lett.} \textbf{\bibinfo{volume}{112}}, \bibinfo{pages}{151803}
  (\bibinfo{year}{2014}), \bibinfo{note}{[Erratum: Phys. Rev.
  Lett.114,no.4,049901(2015)]}, \eprint{1312.2895}.

\bibitem[{\citenamefont{Carli et~al.}(2010)\citenamefont{Carli, Clements,
  Cooper-Sarkar, Gwenlan, Salam, Siegert, Starovoitov, and
  Sutton}}]{Carli:2010rw}
\bibinfo{author}{\bibfnamefont{T.}~\bibnamefont{Carli}},
  \bibinfo{author}{\bibfnamefont{D.}~\bibnamefont{Clements}},
  \bibinfo{author}{\bibfnamefont{A.}~\bibnamefont{Cooper-Sarkar}},
  \bibinfo{author}{\bibfnamefont{C.}~\bibnamefont{Gwenlan}},
  \bibinfo{author}{\bibfnamefont{G.~P.} \bibnamefont{Salam}},
  \bibinfo{author}{\bibfnamefont{F.}~\bibnamefont{Siegert}},
  \bibinfo{author}{\bibfnamefont{P.}~\bibnamefont{Starovoitov}},
  \bibnamefont{and} \bibinfo{author}{\bibfnamefont{M.}~\bibnamefont{Sutton}},
  \bibinfo{journal}{Eur. Phys. J.} \textbf{\bibinfo{volume}{C66}},
  \bibinfo{pages}{503} (\bibinfo{year}{2010}), \eprint{0911.2985}.

\bibitem[{\citenamefont{Catani et~al.}(2010)\citenamefont{Catani, Ferrera, and
  Grazzini}}]{Catani:2010en}
\bibinfo{author}{\bibfnamefont{S.}~\bibnamefont{Catani}},
  \bibinfo{author}{\bibfnamefont{G.}~\bibnamefont{Ferrera}}, \bibnamefont{and}
  \bibinfo{author}{\bibfnamefont{M.}~\bibnamefont{Grazzini}},
  \bibinfo{journal}{JHEP} \textbf{\bibinfo{volume}{05}}, \bibinfo{pages}{006}
  (\bibinfo{year}{2010}), \eprint{1002.3115}.

\bibitem[{\citenamefont{Ball et~al.}(2012)\citenamefont{Ball, Bertone, Cerutti,
  Del~Debbio, Forte, Guffanti, Latorre, Rojo, and Ubiali}}]{Ball:2011uy}
\bibinfo{author}{\bibfnamefont{R.~D.} \bibnamefont{Ball}},
  \bibinfo{author}{\bibfnamefont{V.}~\bibnamefont{Bertone}},
  \bibinfo{author}{\bibfnamefont{F.}~\bibnamefont{Cerutti}},
  \bibinfo{author}{\bibfnamefont{L.}~\bibnamefont{Del~Debbio}},
  \bibinfo{author}{\bibfnamefont{S.}~\bibnamefont{Forte}},
  \bibinfo{author}{\bibfnamefont{A.}~\bibnamefont{Guffanti}},
  \bibinfo{author}{\bibfnamefont{J.~I.} \bibnamefont{Latorre}},
  \bibinfo{author}{\bibfnamefont{J.}~\bibnamefont{Rojo}}, \bibnamefont{and}
  \bibinfo{author}{\bibfnamefont{M.}~\bibnamefont{Ubiali}}
  (\bibinfo{collaboration}{NNPDF}), \bibinfo{journal}{Nucl. Phys.}
  \textbf{\bibinfo{volume}{B855}}, \bibinfo{pages}{153} (\bibinfo{year}{2012}),
  \eprint{1107.2652}.

\bibitem[{\citenamefont{Balazs et~al.}(1995)\citenamefont{Balazs, Qiu, and
  Yuan}}]{Balazs:1995nz}
\bibinfo{author}{\bibfnamefont{C.}~\bibnamefont{Balazs}},
  \bibinfo{author}{\bibfnamefont{J.-w.} \bibnamefont{Qiu}}, \bibnamefont{and}
  \bibinfo{author}{\bibfnamefont{C.~P.} \bibnamefont{Yuan}},
  \bibinfo{journal}{Phys. Lett.} \textbf{\bibinfo{volume}{B355}},
  \bibinfo{pages}{548} (\bibinfo{year}{1995}), \eprint{hep-ph/9505203}.

\bibitem[{\citenamefont{Aliev et~al.}(2011)\citenamefont{Aliev, Lacker,
  Langenfeld, Moch, Uwer et~al.}}]{Aliev:2010zk}
\bibinfo{author}{\bibfnamefont{M.}~\bibnamefont{Aliev}},
  \bibinfo{author}{\bibfnamefont{H.}~\bibnamefont{Lacker}},
  \bibinfo{author}{\bibfnamefont{U.}~\bibnamefont{Langenfeld}},
  \bibinfo{author}{\bibfnamefont{S.}~\bibnamefont{Moch}},
  \bibinfo{author}{\bibfnamefont{P.}~\bibnamefont{Uwer}}, \bibnamefont{et~al.},
  \bibinfo{journal}{Comput.Phys.Commun.} \textbf{\bibinfo{volume}{182}},
  \bibinfo{pages}{1034} (\bibinfo{year}{2011}), \eprint{1007.1327}.

\bibitem[{\citenamefont{Kant et~al.}(2015)\citenamefont{Kant, Kind, Kintscher,
  Lohse, Martini et~al.}}]{Kant:2014oha}
\bibinfo{author}{\bibfnamefont{P.}~\bibnamefont{Kant}},
  \bibinfo{author}{\bibfnamefont{O.}~\bibnamefont{Kind}},
  \bibinfo{author}{\bibfnamefont{T.}~\bibnamefont{Kintscher}},
  \bibinfo{author}{\bibfnamefont{T.}~\bibnamefont{Lohse}},
  \bibinfo{author}{\bibfnamefont{T.}~\bibnamefont{Martini}},
  \bibnamefont{et~al.}, \bibinfo{journal}{Comput.Phys.Commun.}
  \textbf{\bibinfo{volume}{191}}, \bibinfo{pages}{74} (\bibinfo{year}{2015}),
  \eprint{1406.4403}.

\bibitem[{\citenamefont{Harris et~al.}(2002)\citenamefont{Harris, Laenen, Phaf,
  Sullivan, and Weinzierl}}]{Harris:2002md}
\bibinfo{author}{\bibfnamefont{B.}~\bibnamefont{Harris}},
  \bibinfo{author}{\bibfnamefont{E.}~\bibnamefont{Laenen}},
  \bibinfo{author}{\bibfnamefont{L.}~\bibnamefont{Phaf}},
  \bibinfo{author}{\bibfnamefont{Z.}~\bibnamefont{Sullivan}}, \bibnamefont{and}
  \bibinfo{author}{\bibfnamefont{S.}~\bibnamefont{Weinzierl}},
  \bibinfo{journal}{Phys.Rev.} \textbf{\bibinfo{volume}{D66}},
  \bibinfo{pages}{054024} (\bibinfo{year}{2002}), \eprint{hep-ph/0207055}.

\bibitem[{\citenamefont{Campbell et~al.}(2004)\citenamefont{Campbell, Ellis,
  and Tramontano}}]{Campbell:2004ch}
\bibinfo{author}{\bibfnamefont{J.~M.} \bibnamefont{Campbell}},
  \bibinfo{author}{\bibfnamefont{R.~K.} \bibnamefont{Ellis}}, \bibnamefont{and}
  \bibinfo{author}{\bibfnamefont{F.}~\bibnamefont{Tramontano}},
  \bibinfo{journal}{Phys.Rev.} \textbf{\bibinfo{volume}{D70}},
  \bibinfo{pages}{094012} (\bibinfo{year}{2004}), \eprint{hep-ph/0408158}.

\bibitem[{\citenamefont{Langenfeld et~al.}(2009)\citenamefont{Langenfeld, Moch,
  and Uwer}}]{Langenfeld:2009wd}
\bibinfo{author}{\bibfnamefont{U.}~\bibnamefont{Langenfeld}},
  \bibinfo{author}{\bibfnamefont{S.}~\bibnamefont{Moch}}, \bibnamefont{and}
  \bibinfo{author}{\bibfnamefont{P.}~\bibnamefont{Uwer}},
  \bibinfo{journal}{Phys. Rev.} \textbf{\bibinfo{volume}{D80}},
  \bibinfo{pages}{054009} (\bibinfo{year}{2009}), \eprint{0906.5273}.

\bibitem[{\citenamefont{Marquard et~al.}(2015)\citenamefont{Marquard, Smirnov,
  Smirnov, and Steinhauser}}]{Marquard:2015qpa}
\bibinfo{author}{\bibfnamefont{P.}~\bibnamefont{Marquard}},
  \bibinfo{author}{\bibfnamefont{A.~V.} \bibnamefont{Smirnov}},
  \bibinfo{author}{\bibfnamefont{V.~A.} \bibnamefont{Smirnov}},
  \bibnamefont{and}
  \bibinfo{author}{\bibfnamefont{M.}~\bibnamefont{Steinhauser}},
  \bibinfo{journal}{Phys. Rev. Lett.} \textbf{\bibinfo{volume}{114}},
  \bibinfo{pages}{142002} (\bibinfo{year}{2015}), \eprint{1502.01030}.

\bibitem{AbelleiraFernandez:2012cc}
   J.~L.~Abelleira Fernandez {\it et al.} [LHeC Study Group Collaboration],
   %``A Large Hadron Electron Collider at CERN: Report on the Physics and 
Design Concepts for Machine and Detector,''
   J.\ Phys.\ G {\bf 39} (2012) 075001
   [arXiv:1206.2913 [physics.acc-ph]].
   %%CITATION = ARXIV:1206.2913;%%
   %172 citations counted in INSPIRE as of 31 Aug 2015

\end{thebibliography}
%\bibliographystyle{h-physrev5}

}

\newpage


\end{document}